\documentclass[11pt, a4paper]{article}
\usepackage[utf8]{inputenc}
\usepackage[english]{babel}
\usepackage[a4paper,top=2cm,bottom=2.5cm,left=2.5cm,right=2.5cm,marginparwidth=1.75cm]{geometry}
\usepackage{amsmath,amsfonts}
\usepackage{amssymb}
\usepackage{charter}
\usepackage{bbold}
\usepackage{caption,adjustbox}
\usepackage{subcaption}
\usepackage{graphicx}
\usepackage[colorlinks=true,linkcolor=Black,citecolor=Black,urlcolor=Blue]{hyperref}
\usepackage[usenames,dvipsnames]{xcolor}
\usepackage{physics, tensor, cancel}
\usepackage{dsfont}
 \usepackage{float}
\usepackage[vcentermath]{youngtab}
\usepackage{wrapfig}
\usepackage{slashed}
\usepackage{comment}
\usepackage[normalem]{ulem}
\usepackage{fancyhdr}
\usepackage{newtxmath}
\usepackage{calc}        
\usepackage{placeins}
\usepackage{mathtools}
\usepackage{hhline}
\usepackage{makecell}
\usepackage{braket}

\usepackage{booktabs}
\usepackage[toc,page]{appendix}
\usepackage[affil-sl]{authblk}

\title{Four-Fermion Deformations on Line Defects \\in  \\  QED$_4$  with  Yukawas}
\date{}
\author[1,2]{Alessandra D'Alise  \thanks{\href{mailto:alessandra.dalise@unina.it}{alessandra.dalise@unina.it}}}
\author[3]{Giulia Muco \thanks{\href{mailto:giulia@qtc.sdu.dk}{giulia@qtc.sdu.dk}}}
\author[1,2,3,4]{Francesco Sannino\thanks{\href{mailto:sannino@qtc.sdu.dk}{sannino@qtc.sdu.dk}}}

\affil[1]{\small Dept. of Physics E. Pancini, Università di Napoli Federico II, via Cintia, 80126 Napoli, Italy}
\affil[2]{\small INFN sezione di Napoli, via Cintia, 80126 Napoli, Italy}
\affil[3]{\small Quantum  Theory Center ($\hbar$QTC) at IMADA and D-IAS, Southern Denmark Univ., Campusvej 55, 5230 Odense M, Denmark}
\affil[4]{\small Scuola Superiore Meridionale, Largo S. Marcellino, 10, 80138 Napoli NA, Italy}

\begin{document}

\maketitle

\begin{abstract}
We elucidate the dynamics induced by a Wilson Line of charge $q$ and electron charge $e$ in $QED_4$ coupled to a real scalar via Yukawa interactions. We observe that there is a range of the fixed product $e^2 q$ with $e\rightarrow 0$ and $q\rightarrow \infty$ for which four-fermion operators play a dominant role in the dynamics of the theory. We uncover the fixed-point dynamics of the theory and analyze the corresponding critical behavior. 
\end{abstract}

\newpage


\section{Introduction}
Gauge-Yukawa theories constitute the theoretical backbone of the Standard Model of particle interactions. The latter represents, to date, the best description of fundamental interactions up to quantum gravity. The potential dynamical range of these theories is immense spanning from developing a Landau pole to being completely asymptotically free \cite{Gross:1973id,Politzer:1973fx,Callaway:1988ya,Holdom:2014hla,Giudice:2014tma,Pica:2016krb}, or asymptotically safe \cite{Litim:2014uca} in the ultraviolet to featuring infrared free or infrared interacting fixed points \cite{Sannino:2004qp,Dietrich:2006cm,Pica:2016krb}, as well as confining and/or spontaneous symmetry breaking  at low energies. It can also happen that at very high energies the scalars are composite of four-fermion operators (see \cite{Krog:2015bca} for a recent study and references therein).  Many questions remain unanswered because often too hard to tackle either analytically or via numerical methods such as first principle lattice simulations. In this work we inch forward in our understanding of their dynamics by probing it via extended operators such as Wilson lines  \cite{PhysRevD.10.2445}. These correspond to adding infinitely massive charged states to the theory and study their response. Because conformal field theories (CFT)s typically display enlarged symmetries that often appear in either ultraviolet or infrared dynamics they represent a convenient starting point to investigate new dynamics.  We will therefore consider the low energy infrared free conformal limit of four-dimensional QED with a neutral scalar higgs coupled to the dynamical electron via a Yukawa interaction. In fact, in the context of CFTs it is a current matter of interest to study the role of extended operators \cite{Calabrese:2004eu, Billo:2013jda,Cuomo:2022xgw} as reviewed in \cite{Billo:2016cpy}. A possible interest in studying line operators in the Standard Model is the possibility to test the SM quotient gauge group \cite{Tong:2017oea},   $SU(3)\times SU(2)\times U(1)/\Gamma$ (with $\Gamma$ a subgroup of $\mathbf{Z}_6$). 
It is useful to recall that for a line operator on the time
axis (i.e. an impurity fixed in space in the point $\vec{x}=0$), the line operator is said to be conformal if it preserves an $SL(2, \mathbb{R})$ subgroup of the conformal group, defining a  Defect Conformal Field Theory (DCFT).  In literature, examples of interesting line defects in CFTs include symmetry defects \cite{Calabrese:2004eu, Billo:2013jda} and
spin impurities \cite{Cuomo:2022xgw}. When one defines line operators in a CFT, it is not guaranteed that conformality will be preserved. In fact, while the bulk theory stays conformal, typically a renormalization group flow emerges between different line operators, as discussed in \cite{Aharony:2022ntz,Aharony:2023amq}. 
For DCFTs one introduces  the bulk-defect OPE \cite{McAvity:1995zd} that allows to expand a bulk operator $\mathcal{O}_i(\Vec{x},t)$ of scaling dimension $\Delta_i$ in terms of defect operators $\mathcal{U}_{\alpha}(t)$ near the defect (since in our case the defect will be a Wilson Line, we will expand bulk operators around the point $\Vec{x}=\Vec{0}$) as
\begin{equation}
    \label{eq:bulk_def_OPE}
    \mathcal{O}_i(\Vec{x},t)=\sum_{\alpha}\frac{\mathcal{A}_{i\alpha}}{|\Vec{x}|^{\Delta_i-\Delta_{\alpha}}}\mathcal{U}_{\alpha}(t), \ \ \ \ \ \ \ \ \ \ \ |\Vec{x}|\sim0.
\end{equation}
 The associated scaling exponents $\Delta_\alpha$  were computed for the electron field in $QED_4$ \cite{Aharony:2022ntz,Aharony:2023amq}.   Additionally, in absence of the scalar, previous studies \cite{Aharony:2022ntz,Aharony:2023amq} have shown that  $QED_4$  undergoes a phase transition for $\frac{e^2q}{4\pi}>1$. 
 In our case we work in the infrared free regime  where besides taking the same  limits as in \cite{Aharony:2022ntz,Aharony:2023amq} for the electron charge $e\rightarrow 0$ and  the defect charge $q\rightarrow \infty$ with $e^2q$ fixed, we further take the limits of Yukawa $y\to 0$ and scalar self-coupling $\lambda\to 0$ while we further keep fixed the ratios $\lambda/e^2$ and $y/e$. We then study the modified  Renormalization Group dynamics on the Wilson line. We observe that in the dynamical regime $\frac{\sqrt{3}}{2}<\frac{e^2q}{4\pi}<\frac{\sqrt{15}}{4}$ (corresponding to $\frac14<\nu<\frac12$ below), four-fermion deformations play an important role and analyze their impact.

 We determine and analyze the new (unstable) fixed point structure stemming from the emergence of four-fermion operators. The paper is structured as follows. In Section \ref{sec:review}, we revisit key results from scalar and fermionic $QED_4$ in the presence of a Wilson line. Section \ref{sec:QED_Yukawa} introduces the Yukawa interaction, computing the quantum fluctuations in the scalar field using results from the previous section. Section \ref{sec:new_RG} presents our main results on the modified RG flows, including the identification of new fixed points driven by 4-fermion operators. We conclude with a discussion on the physical implications and possible extensions of our analysis.

\section{Introducing the framework: recap of known results}
\label{sec:review}
\subsection{Scalar $QED_4$}
\label{subsec:Scalar_QED4}
In this section we review the results obtained in \cite{Aharony:2022ntz, Aharony:2023amq} when investigating massless scalar $QED$ in $3+1$ dimensions. The insertion of a Wilson line with charge $q$ triggers the response of a given CFT to the presence of this external object. The latter can be written as 
\begin{equation}\label{eq:wilson}
    W(\gamma) = \exp{iq \int_\gamma A_\mu dx^\mu} \ , \qquad q\in\mathbb{Z},
\end{equation}
and it describes the field configuration stemming from an infinitely massive  particle of charge $q$. The action of the full theory in presence of the defect reads\footnote{We work with the metric $g_{\mu\nu}=\text{diag}(1,-1,-1,-1)$ and consider $D_{\mu}=\partial_{\mu}-iA_{\mu}$.} 
\begin{equation}
    \label{eq:ScalarQED}
    S=\int \text{d}^4x\left( -\frac{1}{4e^2}F^{\mu\nu}F_{\mu\nu}+|D_{\mu}\phi|^2-\frac{\lambda}{2}|\phi|^4-q\delta^3(\vec{x})\delta^{\mu}_0A_{\mu}\right) \ .
\end{equation}
 The presence of the defect induces a DCFT which is most suitably studied by rescaling the field $\phi=\Phi/e$ and working in the limit
\begin{equation}
    \label{eq:KomLimit}
    \begin{split}
    &e\to0, \ \ \frac{\lambda}{e^2}=\text{fixed}, \ \ e^2q=\text{fixed},\\
    & \lambda\to0, \ \ q\to\infty   \ .
    \end{split}
\end{equation}
The limit above allows to investigate the dynamics near the non-interacting conformal fixed point via semiclassical approximations. Solving the EOMs in this regime yields
\begin{equation}
    \label{eq:firstSol}
    A_0=\frac{e^2q}{4\pi r}=\frac{g}{r}, \ \ \Phi=0. 
\end{equation}
To determine the fluctuations of $\Phi$ around the defect,   we expand the EOM for small values of $r$ and decompose the field $\Phi$ in Fourier transform and spherical harmonics $Y_{\ell m}(\theta,\varphi)$ as
\begin{equation}
    \label{eq:PhiSol}
    \Phi(t,r,\theta,\varphi)=\sum_{\ell,m}\int d\omega e^{-i\omega t}\Phi_{\omega\ell m}(r)Y_{\ell m}(\theta,\varphi)
\end{equation}
and obtain the following radial dependence
\begin{equation}
    \label{eq:PhiGuess}
    \Phi_{\omega\ell m}\sim r^{\Delta_\pm-1}(1+C_\pm r)\ ,
\end{equation}
where  
\begin{align}
       \Delta_{\pm}=\frac{1}{2}\pm\nu_{\ell}, \qquad
       C_\pm=\mp\frac{g\omega}{\nu_{\ell}\pm\frac{1}{2}},
\end{align}

with $\displaystyle{\nu_{\ell}=\sqrt{\frac{1}{4}+\ell(\ell+1)-g^2}}$ and $\ell\in\mathbb{N}$ is the angular momentum.\\ The generic expression for the field $\Phi$ near the defect is, therefore, a linear combination of the two previous solutions which reads: 
\begin{equation}
\label{eq:PhiSolution}
        \Phi_{\omega\ell m}(r)\sim\alpha_{\omega\ell m}\ r^{-\nu_{\ell}-\frac{1}{2}}\left(1+\frac{g\omega}{\nu_{\ell}-\frac{1}{2}}r+\dots\right)+\beta_{\omega\ell m}\  r^{\nu_{\ell}-\frac{1}{2}}\left(1-\frac{g\omega}{\nu_{\ell}+\frac{1}{2}}r+\dots\right) \ .
\end{equation}
To take into account the impact of the Wilson line one imposes a boundary condition at a small distance $r_0$ from the defect. To do this, we exploit the conformal invariance of the theory by performing a Weyl rescaling to go from $\mathbb{R}^{1,3}$ to $AdS_2\times S^{2}$ space:
\begin{equation}
    \label{eq:Weyl_rescaling_R_to_Ads}
    ds^2=r^2\left(\frac{dt^2-dr^2}{r^2}-d\Omega^2_{2}\right)=r^2d\Tilde{s}^2_{AdS_2\times S^{2}}
\end{equation}
such that now the defect appears at the asymptotic boundary. The bulk action now reads: 
\begin{equation}\label{eq:sbulk}
    S_{\text{bulk}}=\int_{AdS\times S^2}\ \sqrt{-g}\ \mathcal{L}[A,\Phi] \qq{with} \mathcal{L}[A,\Phi]=\frac{1}{e^2}\left[-\frac{1}{4}F_{\mu\nu}F^{\mu\nu}+\abs{D_\mu\Phi}^2-\frac{\lambda}{2e^2}\abs{\Phi}^4\right]  \ ,
\end{equation}
where $\Phi_{AdS_2}=r\Phi$, as a result of the Weyl rescaling. The  conditions at $r_0$ will now emerge from the requirement that our field solution $\Phi$ is an extremal point of our action. Varying the bulk action with respect to the field $\Phi$, we have
\begin{equation}
\label{eq:d_S_bulk}
    \delta S_{\text{bulk}}= \int \sqrt{-g}\ \partial_\mu\left(\fdv{\mathcal{L}}{\partial_\mu \Phi}\ \delta\Phi\right)\ .
\end{equation}
Using the expression in eq. \eqref{eq:PhiSolution} in the above equation, the $\ell=0$ term in the regime of interest $\omega r \ll 1$ is
\begin{equation}
\label{eq:Phi_AdS2}
    \Phi_{AdS_2} = \int\ d\omega\ e^{-i\omega t}\left(\alpha_\omega\ r^{\frac{1}{2}-\nu}+\beta_\omega\ r^{\frac{1}{2}+\nu}\right).
\end{equation}
Plugging this in eq. \eqref{eq:d_S_bulk} implies the following leading term in the integral
\begin{equation}
    \delta S_{\text{bulk}}\ \sim \left(\frac{1-2\nu}{2}\right)\ r_0^{-2\nu}\ \int\ d\omega\ \left(\alpha^\dagger_\omega\ \delta\alpha_\omega+c.c\right)
\end{equation}
indicating that the proper boundary condition to achieve $\delta S_{\text{bulk}}=0$ is
\begin{equation}\label{eq:alphazero}
    \alpha_\omega = \alpha^\dagger_\omega = 0.
\end{equation}
From eq. \eqref{eq:alphazero}, we can read directly the defect scaling dimension of the gauge-invariant operator $\Phi^{\dagger} \Phi$ with this boundary condition using eq. \eqref{eq:bulk_def_OPE}. In fact, the defect scaling dimensions of $\alpha_{\omega}$ and $\beta_{\omega}$ can be read out directly from the powers of $r$ in \eqref{eq:Phi_AdS2}, i.e. $\displaystyle{\hat{\Delta}(\alpha)=\frac{1}{2}-\nu}$ and $\displaystyle{\hat{\Delta}(\beta)=\frac{1}{2}+\nu}$. Hence, $\hat{\Delta}(\Phi^{\dagger}\Phi)=1+2\nu$, which in turn signals that it is irrelevant for $\nu>0$ and marginal for $\nu=0$\footnote{As a consistency check, we stress that for $q=0$ (i.e., for a trivial Wilson line) the defect scaling dimension $\hat{\Delta}_{\Phi^{\dagger}\Phi}$ becomes 2, which coincides with the bulk scaling dimension of the composite operator $\Delta_{\Phi^{\dagger}\Phi}=2$.}. The possibility of this operator to become marginal on the line implies that it is inconsistent to ignore it when studying the dynamics of the theory, and hence we should consider the Wilson Line modified as\footnote{As will be shown, the parameter $f$ here will have a non-trivial RG flow which will be computed with methods similar to \cite{Aharony:2015afa}.}
\begin{equation}
    \label{eq:modified_WL}
    W=\exp\Bigl(iq\int dt A_0(x)-f\int dt \Phi^{\dagger}\Phi\Bigr) \ .
\end{equation}
Modifying the WL in this way amounts to adding boundary terms to our bulk action in eq. \eqref{eq:sbulk}. This will result in no change in our EOMs, but only in changing the boundary conditions on the field $\Phi$. We can for example add to eq. \eqref{eq:sbulk} the term
\begin{equation}\label{eq:sone}
    \mathcal{S}^{(1)}_{\text{bdy}} = -\frac{1-2\nu}{2}\ \int_{r=r_0}\ dt\ \sqrt{-\hat{g}}\ \Phi^\dagger\Phi,
\end{equation}
where the line element on the boundary is $d\hat{s}^2=dt^2/r^2_0$. The term in eq. \eqref{eq:sone}, upon repeating the same analysis as before, yields the boundary condition\footnote{This boundary condition is referred to as the “alternative quantization” of the scalar field in the AdS/CFT literature \cite{Klebanov:1999tb}.} 
\begin{equation}\label{eq:betazero}
    \beta^\dagger_\omega=\beta_\omega= 0
\end{equation}
to cancel out the dominant term in $\delta S_{\text{bulk}}+\delta S^{(1)}_{\text{bdy}}$. The scaling dimension of the gauge-invariant defect operator $\Phi^{\dagger}\Phi$ for the boundary condition in eq. \eqref{eq:betazero} is $\hat{\Delta}_{\alpha^{\dagger}\alpha}=1-2\nu$, meaning it is a relevant operator.  This relevance is significant because adding this operator to the action initiates an RG flow. In fact, including in \eqref{eq:sbulk} both eq. \eqref{eq:sone} and another boundary term
\begin{equation}\label{eq:s2}
    \mathcal{S}^{(2)}_{\text{bdy}} = -f_0 \int_{r_0}\ dt\ \sqrt{-\hat{g}}\ r_0^{2\nu}\ \Phi^\dagger\Phi
\end{equation}
and varying the above boundary term with respect to the fields leads us to
\begin{equation}
   \delta S_{\text{bulk}}+\delta S^{(1)}_{\text{bdy}}+ \delta S^{(2)}_{\text{bdy}} \sim 
   \int\ d\omega\ \bigg(-f_0\ \delta\alpha^\dagger_\omega\alpha_\omega\ +\delta\alpha^\dagger_\omega\beta_\omega (2\nu-f_0 r_0^{2\nu})+\dots\bigg)
\end{equation}
which is zero if 
\begin{equation}
    \frac{\beta_\omega}{\alpha_\omega} = \frac{f}{2\nu-f}\ r_0^{-2\nu}\qq{where} f=f_0 r_0^{2\nu}
\end{equation}
is satisfied. From the above equation we understand that the coupling $f$ runs. To write down the beta function associated to it we first require that the above boundary condition does not depend on the cut-off, i.e. the unphysical scale $\mu=r_0^{-1}$, then  
\begin{equation}
    \pdv{}{\log\mu}\left(\frac{\beta_\omega}{\alpha_\omega}\right)=0\implies \pdv{}{f}\left(\frac{\beta_\omega}{\alpha_\omega}\right)\beta_f+\pdv{}{r_0}\left(\frac{\beta_\omega}{\alpha_\omega}\right)\pdv{r_0}{\log\mu}= 0
\end{equation}
yields \cite{Aharony:2023amq}
\begin{equation}
   \beta_f = f(f-2\nu)\ , 
\end{equation}
meaning that the $\beta_f$ admits two zeroes: one at $f=0$ (UV fixed point) and the other at $f=2\nu$ (IR fixed point). This RG flow actually corresponds to interpolating between the boundary conditions for the field $\Phi$ in eq. \eqref{eq:alphazero} and eq. \eqref{eq:betazero}.
 This flow is initiated by the gauge-invariant relevant defect operator $|\Phi|^2$, which has a dimension of $1 - 2\nu$ in the UV DCFT and becomes irrelevant with a dimension of $1 + 2\nu$ in the IR DCFT. Although the charge of the Wilson line is quantized and does not flow under the RG, other couplings localized on the defect related to the additional fields in the theory do flow non-trivially.


\subsection{Fermionic $QED_4$}
\label{subses:FermionicQED}
Let us now focus on the example of fermionic QED in four space-time dimensions with metric $\eta_{\mu\nu}=\text{diag}(1,-\vec{1})$. In the presence of a Wilson Line of charge $q$ fixed in space at the position $\vec{x}=0$, the Lagrangian of the theory is
\begin{equation}
    \label{eq:Femrionic_QED_L}
    \mathcal{L}=-\frac{1}{4e^2}F_{\mu\nu}F^{\mu\nu}+i\bar{\psi}\slashed{D}\psi-q\delta^3(\vec{x})\delta^{\mu}_0A_{\mu}(x)
\end{equation}
with $F_{\mu\nu}$ the usual electromagnetic strength tensor, $D_{\mu}=\partial_{\mu}-iA_{\mu}$ being the covariant derivative and $\psi$ being a massless Dirac spinor with charge 1 under the $U(1)$ gauge symmetry. Let us rescale the spinor field as $\displaystyle{\psi=\frac{\Psi}{e}}$, so that the action of the theory becomes
\begin{equation}
    \label{eq:Femrionic_QED_S}
    \mathcal{S}=\frac{1}{e^2}\int d^4 x\Bigl[ -\frac{1}{4}F_{\mu\nu}F^{\mu\nu}+i\bar{\Psi}\slashed{D}\Psi-e^2q\delta^3(\vec{x})\delta^{\mu}_0A_{\mu}(x) \Bigr]
\end{equation}
As discussed in the previous section, we can now take the double scaling limit
\begin{equation}
    \label{eq:double_scaling}
        e\to0, \ \ \ q\to\infty, \ \ \ e^2 q=\text{fixed}
\end{equation}
so that the partition function $\displaystyle{\mathcal{Z}=\int \mathcal{D}A\mathcal{D}\Psi \exp{-i\mathcal{S}[A,\Psi]}}$ of the theory is dominated by the minima of $\mathcal{S}[A,\Psi]$, allowing us to work in a semiclassical approximation. Moreover, in such a limit we are placing ourselves in a zero of the beta function $\beta(e)$, making the theory conformal also at the quantum level, allowing us later to exploit the Weyl invariance of the theory in our computations.\\
As before, our goal is to write the spinor field operator near the defect (i.e. for $r\sim0$) in terms of defect operators, whose scaling dimension will be given by the bulk-defect OPE in eq. \eqref{eq:bulk_def_OPE}.\\
We hence write the EOMs
\begin{equation}
    \label{eq:EOMs_fermion_QED}
    \begin{cases}
        \partial_{\mu}F^{\mu\nu}=-j^{\nu}+e^2q\delta^2(\vec{x})\delta^{\nu}_0\\
        i\slashed{D}\Psi =0
    \end{cases},
\end{equation}
where $j^{\nu}=\bar{\Psi}\Gamma^{\nu}\Psi$. As a first approximation, we can take the classical solution to the eq.s \eqref{eq:EOMs_fermion_QED} to be
\begin{equation}
    \label{eq:ferm_QED_class_sol}
    A_0(x)=\frac{g}{r}, \ \ \Psi=0,
\end{equation}
with
\begin{equation}
\label{eq:coupling_def}
     g=\frac{e^2q}{4\pi}.
\end{equation}
Plugging the solution for $A_{\mu}$ back into the EOM for $\Psi$, we find its quantum fluctuations in the presence of the Coulomb background of eq. \eqref{eq:ferm_QED_class_sol}. To do this, as in subsection \ref{subsec:Scalar_QED4}, we perform a Weyl rescaling of the metric to go from flat space to $AdS_2\times S^2$ and decompose the spinor field as \cite{Lopez-Ortega:2009flo}
\begin{equation}
    \label{eq:Fermion_expansion}
    \Psi=\frac{1}{r^{\frac{3}{2}}}\sum_{\ell, s}\sum_{\delta=+,-}\psi^{(\delta)}_{\ell s}(t,r)\otimes \chi^{(\delta)}_{\ell s}(\hat{n})
\end{equation}
By doing this, with the steps shown in Appendix \ref{subsec:details_fermion}, we obtain that the action for a $d$-dimensional Dirac spinor in the presence of a $U(1)$ gauge field can be decomposed into the sum of actions for $AdS_2$ spinors with masses $m_{\ell}$, i.e.
\begin{equation}
    \label{eq:final_ferm_S}
    \mathcal{S}=\sum_{\ell s}\sum_{\delta=\pm}\int_{AdS_2}d^2x \sqrt{g}\bar{\psi}^{(\delta)}_{\ell s}[i(\slashed{\nabla}_{AdS_2}-i\slashed{A})-m_{\ell}]\psi^{(\delta)}_{\ell s},
\end{equation}
where $m_{\ell}=\displaystyle{\ell+1}$ in $d=4$ dimensions.\\
Motivated by this, in the following we will study a single Dirac fermion in $AdS_2$ in the presence of a background field $A_0=g/r$. This will allow us to quickly discuss the case of a Dirac fermion in $d=4$. So, we study the action\footnote{This action is written in such a way that it is exactly hermitian, and not only up to surface terms.}
\begin{equation}
    \label{eq:spinors_AdS2}
    \mathcal{S}=\int_{AdS_2}d^2 x \sqrt{g}\bar{\psi}\Bigl[i\Bigl(\overset\leftrightarrow{\slashed{\nabla}}_{AdS_2}-i\slashed{A}\Bigr)-m_{\ell}\Bigr]\psi,
\end{equation}
where $\slashed{\nabla}_{AdS_2}=\gamma^a e^{ \ \ \mu}_{a}\nabla_{\mu}$, $e^{\mu}_{ \ a}$ being the vielbeins and $\nabla_{\mu}=\displaystyle{\partial_{\mu}+\frac{1}{8}\omega_{\mu ab}[\gamma^a,\gamma^b]}$ the spinor covariant derivative, and $\slashed{A}=\gamma^a e^{\mu}_{ \ a}A_{\mu}$. If we choose the representation for the Clifford Algebra to be $\gamma^0=\sigma^1, \ \gamma^1=i\sigma^3, \ \gamma^3=\gamma^0\gamma^1$,
we get the equation of motion for $\psi_{\ell}$, which is (neglecting the time dependence)
\begin{equation}
    \label{eq:psi_eom}
    \left[i\left(r\gamma^1\partial_r-\frac{1}{2}\gamma^1-ig\gamma^0\right)-m_{\ell} \right]\psi_{\ell}(r)=0.
\end{equation}
The steps that lead to this equation are shown in Appendix \ref{app:AdSS2_fermions}.\\
So, writing
\begin{equation}
    \label{eq:spinor_AdS}
    \psi_{\ell}=\begin{pmatrix}
        \zeta_{\ell} \\
        \xi_{\ell}
    \end{pmatrix}
\end{equation}
with $\zeta$ and $\xi$ both single-component Grassmann fields, and looking for solutions of the form $\sim r^{\Delta}$ near $r\sim0$ for both of them, we find 
\begin{equation}
    \label{eq:Delta_fermions}
    \Delta_{\pm}=\frac{1}{2}\pm \nu_{\ell},
\end{equation}
hence
\begin{equation}
    \label{eq:psi_sol}
    \begin{split}
        &\zeta_{\ell} = \alpha_{\ell} r^{\frac{1}{2}-\nu_{\ell}} + \frac{g}{m_{\ell}+\nu_{\ell}}\beta_{\ell}r^{\frac{1}{2}+\nu_{\ell}}\\
        &\xi_{\ell} = \beta_{\ell}r^{\frac{1}{2}+\nu_{\ell}}+\frac{g}{m_{\ell}+\nu_{\ell}}\alpha_{\ell}r^{\frac{1}{2}-\nu_{\ell}}
    \end{split}
\end{equation}
with 
\begin{equation}
    \label{eq:nu_ell_fermion}
    \nu_{\ell}=\sqrt{m^2_{\ell}-g^2}.
\end{equation}
We now carry on the discussion as in subsection \ref{subsec:Scalar_QED4}, meaning that we will add boundary terms to the action in eq. \eqref{eq:spinors_AdS2} in order to get two boundary conditions which are fixed points of an RG flow triggered by relevant operators in the theory. Because of the unitarity bound $\Delta>0$, from eq. \eqref{eq:Delta_fermions} we will work in the interval $0<\nu_{\ell}<1/2$. Moreover, remembering that the action we are originally interested in is the one in eq. \eqref{eq:final_ferm_S}, we will focus on the mode with angular momentum $\ell=0$, as it has the lightest $AdS_2$ mass and for simplicity we will hence drop the subscript $\ell$ in the following. With this being said, varying the bulk action as in eq. \eqref{eq:d_S_bulk}, we obtain that the field configurations that minimize the bulk action must make equal to zero the expression
\begin{equation}
 \label{eq:d_S_ferm}
    \delta S_{\text{bulk}}=-\frac{i}{2}\int_{r_0}dt\sqrt{g g^{rr}}(\Bar{\psi}\gamma^1\delta\psi-\delta\Bar{\psi}\gamma^1\psi)
\end{equation}
which, using the expressions in eq. \eqref{eq:psi_sol}, becomes
\begin{equation}
    \label{eq:d_S_ferm_2}
        \delta S_{\text{bulk}}=\frac{\nu}{m+\nu}\int_{r_0} dt(\beta^{\dagger}\delta\alpha-\alpha^{\dagger}\delta\beta+\delta\alpha^{\dagger}\beta-\delta\beta^{\dagger}\alpha).
\end{equation}
It is technically convenient now to notice that the possible boundary terms are any linear combination of the bilinears in the fermion field\footnote{Again, we only consider bilinears as we need to consider gauge-invariant quantities.} such as $\bar{\Psi}\Psi, \ \bar{\Psi}\gamma^0\Psi, \ \bar{\Psi}\gamma^1\Psi, \ \bar{\Psi}\gamma^3\Psi$. In terms of the boundary modes $\alpha$ and $\beta$, this is equivalent to considering any linear combination of the bilinears\footnote{In the computation the determinant of the metric on the boundary is already considered.} $\beta^{\dagger}\beta r_0^{2\nu}, \ \beta^{\dagger}\alpha, \ \alpha^{\dagger}\beta, \ \alpha^{\dagger}\alpha r^{-2\nu}_0$.\\
With this being said, a possible boundary term to add to the action \eqref{eq:final_ferm_S} is
\begin{equation}
    \label{eq:ferm_bdy_1}
    \mathcal{S}^{(1)}_{\text{bdy}}=\frac{\nu}{m+\nu}\int_{r_0}dt (\beta^{\dagger}\alpha+\alpha^{\dagger}\beta+2\beta^{\dagger}\beta r_0^{2\nu})
\end{equation}
In fact,
\begin{equation}
    \label{eq:d_ferm_bdy_1}
   \delta \mathcal{S}^{(1)}_{\text{bdy}}=\frac{\nu}{m+\nu}\int_{r_0}dt [\delta\beta^{\dagger}\alpha+\beta^{\dagger}\delta\alpha+\delta\alpha^{\dagger}\beta+\alpha^{\dagger}\delta\beta+2(\delta\beta^{\dagger}\beta +\beta^{\dagger}\delta\beta )r_0^{2\nu}]
\end{equation}
Hence, the condition
\begin{equation}
    \delta(\mathcal{S}_{\text{bulk}}+ \mathcal{S}^{(1)}_{\text{bdy}})=0
\end{equation}
yields
\begin{equation}
    \frac{2\nu}{m+\nu}\int_{r_0}dt [\beta^{\dagger}\delta\alpha+\delta\alpha^{\dagger}\beta+2(\delta\beta^{\dagger}\beta +\beta^{\dagger}\delta\beta )r_0^{2\nu}]=0
\end{equation}
which implies the boundary condition
\begin{equation}
\label{eq:bdy_cond_ferm_1}
    \beta=\beta^{\dagger}=0,
\end{equation}
while the boundary modes $\alpha$ and $\alpha^{\dagger}$ are left free to fluctuate. An alternative boundary term is
\begin{equation}
    \label{eq:ferm_bdy_2}
    \mathcal{S}^{(2)}_{\text{bdy}}=-\frac{\nu}{m+\nu}\int_{r_0}dt (\beta^{\dagger}\alpha+\alpha^{\dagger}\beta+2\alpha^{\dagger}\alpha r_0^{-2\nu})
\end{equation}
With this, the condition
\begin{equation}
      \delta(\mathcal{S}_{\text{bulk}}+ \mathcal{S}^{(2)}_{\text{bdy}})=0
\end{equation}
yields the boundary condition
\begin{equation}
\label{eq:bdy_cond_ferm_2}
    \alpha=\alpha^{\dagger}=0,
\end{equation}
while the modes $\beta$ and $\beta^{\dagger}$ are free to fluctuate.\\
Finally, we want to show that the two boundary conditions \eqref{eq:bdy_cond_ferm_1} and \eqref{eq:bdy_cond_ferm_2} correspond to the fixed points of an RG flow. In fact, as in subsection \ref{subsec:Scalar_QED4}, there appears in the theory an operator which is relevant on the line, which in this case is the double-trace operator $\alpha^{\dagger}\alpha$, whose scaling dimension is $\hat{\Delta}(\alpha^{\dagger}\alpha)=1-2\nu<1$. To see this RG flow, let us deform the Wilson Line with the boundary operator
\begin{equation}
    \label{eq:S_DTD_ferm}
    \mathcal{S}^{DTD}_{\text{bdy}}=-2f_0\int_{r_0}dt r_0^{2\nu}(\beta^{\dagger}\alpha+\alpha^{\dagger}\beta+\alpha^{\dagger}\alpha r_0^{-2\nu}+\beta^{\dagger}\beta r_0^{2\nu}),
\end{equation}
which is such that $\displaystyle{\mathcal{S}^{DTD}_{\text{bdy}}\xrightarrow[]{r_0\to0}-2f_0\int dt \alpha^{\dagger}\alpha }$, with $f_0$ a dimensionful bare coupling. Hence, by requiring
\begin{equation}
    \delta( \mathcal{S}_{\text{bulk}}  +\mathcal{S}^{(1)}_{\text{bdy}}+\mathcal{S}^{DTD}_{\text{bdy}})=0
\end{equation}
we get the boundary condition
\begin{equation}
\label{eq:bound_cond_ferm_3}
    \frac{\beta}{\alpha}=\frac{f_0 r_0^{2\nu}(m+\nu)}{\nu-f_0 r_0^{2\nu}(m+\nu)}r_0^{-2\nu}.
\end{equation}
From here, requiring that the boundary conditions do not depend on the arbitrary scale $\displaystyle{\mu=\frac{1}{r_0}}$, we can write the  Callan-Symanzik equation
\begin{equation}
    \label{eq:Callan_Symanzik_ferm}
    \frac{\partial}{\partial\log{\mu}}\Bigl(\frac{\beta}{\alpha}\Bigr)=0 \ \ \Rightarrow \ \ \frac{\partial}{\partial f}\Bigl(\frac{\beta}{\alpha}\Bigr) \beta_f -r_0\frac{\partial}{\partial r_0}\Bigl(\frac{\beta}{\alpha}\Bigr)=0
\end{equation}
where we have defined the dimensionless parameter $f=f_0 r_0^{2\nu}$ and $\displaystyle{\beta_f=  \frac{\partial}{\partial\log{\mu}}f}$ is its beta function, which we find by solving eq. \eqref{eq:Callan_Symanzik_ferm}. Thus, we find
\begin{equation}
    \label{eq:beta_f_ferm}
    \beta_f=-2f[\nu-f(m+\nu)].
\end{equation}
As anticipated, this beta function has two fixed points, each corresponding to one of the boundary conditions \eqref{eq:bdy_cond_ferm_1} and \eqref{eq:bdy_cond_ferm_2}, as can be seen from eq. \eqref{eq:bound_cond_ferm_3}:
\begin{equation}
\label{eq:ferm_fixed_points}
\begin{cases}
    f=0 \  \ \ \     \ \ \ \Rightarrow \ \ \ \beta=0\\
    f=\frac{\nu}{m+\nu}  \ \ \ \Rightarrow \ \ \ \alpha=0
\end{cases},
\end{equation}
the first one corresponding to a UV unstable fixed point (where there is, in fact, present the relevant operator $\alpha^{\dagger}\alpha$) and the second fixed point corresponding to an IR stable DCFT.\\
As anticipated, we can apply this analysis of a Dirac fermion in $AdS_2$ to the case of a Dirac fermion in $d=4$. In order to do this, we restore the indices $\psi\to\psi^{(\delta)}_s$ and define the $\displaystyle{\psi^{(\delta)}=\Big(\psi^{(\delta)}_{\frac{1}{2}} \ \ \psi^{(\delta)}_{-\frac{1}{2}} \Big)^T}$ vectors, which are doublets of $SU(2)$. Then, the most general gauge-invariant bilinear we can add to the action in eq. \eqref{eq:final_ferm_S} which preserves its $SU(2)$ invariance will be a linear combination of the terms\footnote{It is possible to generalise this analysis further and contemplate non-$SU(2)$ invariant bilinears. We refer the reader to reference \cite{Aharony:2023amq} for this.}
\begin{equation}
    \label{eq:ferm_4d_bilinear}
    \Phi^{(\delta\gamma)}\coloneqq r_0^{2\nu}(\beta^{(\delta)\dagger}\beta^{(\gamma)}r_0^{2\nu}+\beta^{(\delta)\dagger}\alpha^{(\gamma)}+\alpha^{(\delta)\dagger}\beta^{(\gamma)}+r_0^{-2\nu}\alpha^{(\delta)\dagger}\alpha^{(\gamma)}),
\end{equation}
giving in total four combinations, which can be written as $f_0( \Phi^{(++)}+ \Phi^{(--)})+ik_0 ( \Phi^{(+-)}- \Phi^{(-+)})+h_0 ( \Phi^{(+-)}+ \Phi^{(-+)})+q_0( \Phi^{(++)}- \Phi^{(--)})$ (so that the four couplings $f_0, \ k_0, \ h_0, \ q_0$ are real). At this point, we perturb the action \eqref{eq:final_ferm_S} by adding the boundary term
\begin{equation}
    \label{eq:bdy_fermion_4d}
    \mathcal{S}_{\text{DTD}}=-2r_0^{2\nu}\int_{r_0}dt(\beta^{\dagger}F_0\beta r_0^{2\nu}+\beta^{\dagger}F_0\alpha+\alpha^{\dagger}F_0\beta+r_0^{-2\nu}\alpha^{\dagger}F_0\alpha),
\end{equation}
where we have defined $\displaystyle{\alpha\coloneqq \Big(\alpha^{(+)}_{\frac{1}{2}} \ \alpha^{(+)}_{-\frac{1}{2}} \ \alpha^{(-)}_{\frac{1}{2}} \ \alpha^{(-)}_{-\frac{1}{2}}\Big)^T}$ ($\beta$ analogously) and $F_0$ is the $4\times4$ hermitian matrix $\displaystyle{F_0\coloneqq\lambda_A\Sigma^A}$ with $\lambda^A=f_0, \ k_0, \ h_0, \ q_0$ and
\begin{equation}
    \label{eq:sigmas}
    \Sigma^{f_0}=\begin{pmatrix}
        \mathds{1}_2 & 0 \\
        0 & \mathds{1}_2
    \end{pmatrix}, \ \  \Sigma^{k_0}=\begin{pmatrix}
        0 & i\mathds{1}_2 \\
        -i\mathds{1}_2 & 0
    \end{pmatrix}, \ \ \Sigma^{h_0}=\begin{pmatrix}
        0 & \mathds{1}_2\\
        \mathds{1}_2 & 0
    \end{pmatrix}, \ \ \Sigma^{q_0}=\begin{pmatrix}
        \mathds{1}_2 & 0 \\
        0 & -\mathds{1}_2
    \end{pmatrix},
\end{equation}
$\mathds{1}_2$ being the $2\times 2 $ identity matrix. The coupling constants are then related to $F_0$ via
\begin{equation}
    \lambda_A=\frac{1}{4}\text{Tr}(\Sigma^AF_0).
\end{equation}
Carrying out the analysis similarly to before, requiring that the variation of the total action is zero, the emerging boundary conditions are
\begin{equation}
    \beta=C\alpha, \ \ C=(m+\nu)F_0\cdot[\nu\mathds{1}_4-(m+\nu)F_0r_0^{2\nu}]^{-1},
\end{equation}
where $\mathds{1}_4$ is the $4\times4$ identity matrix. Imposing the Callan-Symanzik equation, we find the beta function
\begin{equation}
    \beta_F=-2\nu F+2(m+\nu)F\cdot F \ ,
\end{equation}
with $F=F_0r_0^{2\nu}$. The beta functions of the single couplings are then obtained via
\begin{equation}
    \beta_{\lambda_A}=\frac{1}{4}\text{Tr}(\Sigma^A\beta_F),
\end{equation}
resulting in
\begin{equation}
    \begin{split}
        &\beta_{f}=-2\nu f +2(m+\nu)[f^2+k^2+h^2+q^2],\\
        &\beta_{k}=-2\nu k +4(m+\nu)kf,\\
        &\beta_{h}=-2\nu h +4(m+\nu)hf,\\
        &\beta_{q}=-2\nu q +4(m+\nu)qf.\\
    \end{split}
\end{equation}
The fixed points can be finally classified in the following way:
\begin{itemize}
    \item $(f,k,h,q)=(0,0,0,0)$: an unstable fixed point where the $\beta$ modes are zero;
    \item $(f,k,h,q)=\Big(\frac{\nu}{m+\nu},0,0,0\Big)$: a stable fixed point that corresponds to the $\alpha$ modes being zero;
    \item $f=\frac{\nu}{2(m+\nu)}$ while $k^2+h^2+q^2=\frac{\nu^2}{4(m+\nu)^2}$, which is a set of unstable fixed points corresponding to mixed boundary conditions.
\end{itemize}

\section{Quartic deformations in scalar QED$_4$}
\begin{figure}[h!]
\begin{center}
\includegraphics[width=9cm]{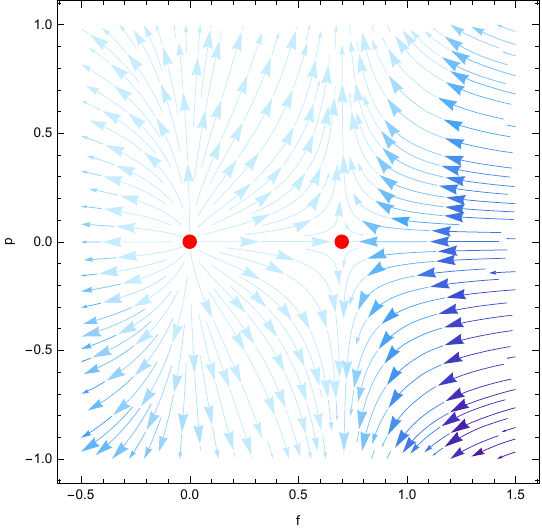}
\end{center}
\caption{\label{fig:phasediagram}Phase diagram of Scalar QED$_4$ with quadratic and quartic defect deformations. In red, the fixed points of the theory. Since $\displaystyle{1/4<\nu<1/2}$, here we fixed $\nu=0.35$. The term $-p\beta_f/(\nu+f)$ in the corrected $\beta_p$ vanishes at both fixed points and is omitted from the plot.}
\end{figure}

As in the next sections it will be necessary to treat quartic deformations on the defect, it is convenient to first study them in scalar QED$_4$ as a warm-up. This is the goal in this section.\\
As in Section \ref{subsec:Scalar_QED4}, our theory is
\begin{equation}
    \begin{split}
    \mathcal{S}_{\text{bulk}}=&\int \text{d}^4x\left( -\frac{1}{4e^2}F^{\mu\nu}F_{\mu\nu}+|D_{\mu}\phi|^2-\frac{\lambda}{2}|\phi|^4-q\delta^3(\vec{x})\delta^{\mu}_0A_{\mu}\right)=\\
        =&\frac{1}{e^2}\int \text{d}^4x\left( -\frac{1}{4}F^{\mu\nu}F_{\mu\nu}+|D_{\mu}\Phi|^2-\frac{\lambda}{2e^2}|\Phi|^4-e^2q\delta^3(\vec{x})\delta^{\mu}_0A_{\mu}\right)
    \end{split}
\end{equation}
Now, we remind the reader that decomposing the field in modes as
\begin{equation}
    \Phi(t,r,\theta,\varphi)=\sum_{\ell,m}\int d\omega e^{-i\omega t}\phi_{\omega\ell m}(r)Y_{\ell m}(\theta,\varphi)=\sum_{\ell,m}\int d\omega \phi_{\omega}(t,r,\theta,\varphi)
\end{equation}
we get the following solution for the radial part
\begin{equation}
\label{eq:radial_phi_quartic}
        \phi_A(r)\sim\alpha_A r^{-\nu_{\ell}-\frac{1}{2}}+\beta_A r^{\nu_{\ell}-\frac{1}{2}} \ ,
\end{equation}
and we have a variation of the bulk action
\begin{equation}
    \delta \mathcal{S}_{\text{bulk}}\ \sim \left(\frac{1-2\nu}{2}\right)\ r_0^{-2\nu}\ \int\ d\omega\ \left(\alpha^\dagger_\omega\ \delta\alpha_\omega+c.c\right) \ .
\end{equation}
Now, as discussed in the previous sections, the quadratic deformations $\alpha^2$ have scaling dimension $\Delta(\alpha^2)=1-2\nu$, and are thus relevant in the range $\nu<1/2$. However, the study in Section \ref{subsec:Scalar_QED4} is not totally complete: in fact, we can also write the gauge-invariant operator $\alpha^4$, whose scaling dimension is $\Delta(\alpha^4)=2-4\nu$, which becomes relevant for $\displaystyle{\nu>\frac{1}{4}}$. So, to push the analysis, for the rest of this section we focus on the range $\displaystyle{\frac{1}{2}>\nu>\frac{1}{4}}$.\\
If we try to deform our theory with a quartic interaction from here, we have
\begin{equation}
\begin{split}
\int_{r_0} dt\,\sqrt{\hat g}\,|\Phi|^4
&=
\sum_{\substack{\ell,\ell',\ell'',\ell'''\\
m,m',m'',m'''}}
\int_{r_0} dt\,
\frac{1}{r_0}
\int d\omega\,d\omega'\,d\omega''\,d\omega''' \\
&\qquad\times
e^{-i[(\omega-\omega')+(\omega''-\omega''')]t}
\,
\phi_A\phi_{A'}\phi_{A''}\phi_{A'''} \\
&=
\sum_{\substack{\ell,\ell',\ell'',\ell'''\\
m,m',m'',m'''}}
\frac{2\pi}{r_0}
\int d\omega\,d\omega'\,d\omega''\,d\omega'''
\,
\delta\!\left[(\omega-\omega')+(\omega''-\omega''')\right]
\phi_A\phi_{A'}\phi_{A''}\phi_{A'''} \ .
\end{split}
\end{equation}
In this way, it is not possible to bring this into a single integral in the $\omega$ variable like in $\delta \mathcal{S}_{\text{bulk}}$ above. To proceed, we observe that we can neglect the time dependence of $\Phi$, i.e.
\begin{equation}
 \Phi(t,r,\theta,\varphi)=\sum_{\ell,m}\phi_{\ell m}(r)Y_{\ell m}(\theta,\varphi) \ .
\end{equation}
It can be checked that this does not vary the solution for the radial part, which stays the one in eq. \eqref{eq:radial_phi_quartic}, and the variation of the bulk action becomes
\begin{equation}
    \delta \mathcal{S}_{\text{bulk}}\ \sim \left(\frac{1-2\nu}{2}\right)\ r_0^{-2\nu}\ \int\ dt \left(\alpha^\dagger\ \delta\alpha+c.c\right) \ .
\end{equation}
Then, we can again take the two quadratic deformations
\begin{equation}
\begin{split}
     \mathcal{S}^{(1)}_{\text{bdy}} &= -\frac{1-2\nu}{2}\ \int_{r=r_0}\ dt\ \sqrt{-\hat{g}}\ \Phi^\dagger\Phi \ ,\\
      \mathcal{S}^{(2)}_{\text{bdy}} &= -f_0 \int_{r_0}\ dt\ \sqrt{-\hat{g}}\ r_0^{2\nu}\ \Phi^\dagger\Phi \ 
\end{split}
\end{equation}
and further add the quartic one
\begin{equation}
    \mathcal{S}^{(3)}_{\text{bdy}}=r_0^{4\nu-1}p_0\int_{r_0} dt \sqrt{-\hat{g}}(\Phi^{\dagger}\Phi)^2 \ \xrightarrow[r_0\to0]{} \ p_0\int dt (\alpha^{\dagger}\alpha)^2 \ ,
\end{equation}
with both $f_0$ and $p_0$ dimensionful couplings that are expected to produce an RG flow in the theory. We now find their beta functions by finding boundary conditions between $\alpha$ and $\beta$ and demanding that they are independent of the unphysical cut-off scale $r_0$. We have then
\begin{align}
\delta\!\left(
\mathcal{S}_{\mathrm{bulk}}
+  \mathcal{S}_{\mathrm{bdy}}^{(1)}
+   \mathcal{S}_{\mathrm{bdy}}^{(2)}
+ \mathcal{S}_{\mathrm{bdy}}^{(3)}
\right)
&=
\int dt\,
\Bigg\{
\delta\alpha^\dagger
\Big[
-f_0 \alpha
+ \beta\left(2\nu-f_0 r_0^{2\nu}\right)
\nonumber\\
&\qquad
+p_0\Big(
\alpha\alpha^\dagger\alpha
+\beta\,\alpha^\dagger\alpha\,r_0^{2\nu}
+\alpha\beta^\dagger\beta\,r_0^{4\nu}
+\beta\beta^\dagger\beta\,r_0^{6\nu}
\nonumber\\
&\qquad\qquad
+\alpha\alpha^\dagger\beta\,r_0^{2\nu}
+\alpha\beta^\dagger\alpha\,r_0^{2\nu}
+\beta\alpha^\dagger\beta\,r_0^{4\nu}
+\beta\beta^\dagger\alpha\,r_0^{4\nu}
\Big)
\Big]
\nonumber\\
&\qquad
+\delta\beta^\dagger
\left(
\text{same coefficient}
\right)
+\delta\mathrm{sub.}
\Bigg\}
=0 \ ,
\end{align}
which gives the boundary conditions
\begin{equation}
-f_0 \alpha
+\beta\left(2\nu-f_0 r_0^{2\nu}\right)
+p_0\!\left(\alpha
\alpha^\dagger\alpha
+\beta\alpha^\dagger\alpha\,r_0^{2\nu}
+\alpha\alpha^\dagger\beta\,r_0^{2\nu}
+\alpha\beta^\dagger\alpha\,r_0^{2\nu}
\right)
=0.
\end{equation}
and
\begin{equation}
-f_0 \alpha^{\dagger}
+\beta^{\dagger}\left(2\nu-f_0 r_0^{2\nu}\right)
+p_0\!\left(\alpha^{\dagger}
\alpha^\dagger\alpha
+2\beta^{\dagger}\alpha^\dagger\alpha\,r_0^{2\nu}
+\alpha^{\dagger} \alpha^{\dagger}\beta\,r_0^{2\nu}
\right)
=0 \ .
\end{equation}
So,
\begin{equation}
\beta
\left(
2\nu
-f_0 r_0^{2\nu}
+2r_0^{2\nu}p_0\,\alpha^\dagger\alpha
\right)
=
f_0 \alpha
-p_0 (\alpha\alpha^\dagger\alpha +r_0^{2\nu}\alpha\beta^{\dagger}\alpha) \ .
\end{equation}
From the second boundary condition we find $\beta^{\dagger}$ as
\begin{equation}
    \beta^{\dagger}=\frac{f_0\alpha^{\dagger}-p_0(\alpha^{\dagger}\alpha^{\dagger}\alpha+r_0^{2\nu}\alpha^{\dagger}\alpha^{\dagger}\beta)}{2\nu\Big(1-\frac{f_0r_0^{2\nu}-2p_0r_0^{2\nu}\alpha^{\dagger}\alpha}{2\nu}\Big)} \ ,
\end{equation}
which can be substituted into the equation for $\beta(\alpha)$. Using $\displaystyle{\frac{1}{1-x}\simeq 1+x}$ and only keeping linear orders in $p$, we finally have the boundary condition 
\begin{equation}
\label{eq:scalar_bdy_cond_final}
        \beta
=\frac{f r_0^{-2\nu}}{2\nu}\Bigg(1 +
\frac{f }{2\nu}\Bigg)\alpha-r_0^{1-4\nu}p\Bigg(1+\frac{f}{\nu}\Bigg) \alpha \alpha^\dagger \alpha \ .
\end{equation}
Like this, demanding that equation \eqref{eq:scalar_bdy_cond_final} does not depend on $\mu=r_0^{-1}$, we find the two beta functions\footnote{The derivation of $\beta_p$ proceeds analogously to the one for $\beta_f$. We report it here step by step. Writing the boundary condition \eqref{eq:scalar_bdy_cond_final} as $\beta=c_1\,\alpha+c_2\,\alpha\alpha^\dagger\alpha$ with
\begin{equation}
 c_1=\frac{f\,r_0^{-2\nu}}{2\nu}\Big(1+\frac{f}{2\nu}\Big)\ ,\qquad c_2=-r_0^{1-4\nu}\,p\,\Big(1+\frac{f}{\nu}\Big)\ ,
\end{equation}
we require the quartic coefficient $c_2$ to be independent of $\mu=r_0^{-1}$. Using $\frac{d\,r_0^{\,k}}{d\log\mu}=-k\,r_0^{\,k}$, we obtain
\begin{equation}
\begin{split}
 &0=\frac{d c_2}{d\log\mu}=-r_0^{1-4\nu}\left[\beta_p\Big(1+\frac{f}{\nu}\Big)+\frac{p\,\beta_f}{\nu}-(1-4\nu)\,p\,\Big(1+\frac{f}{\nu}\Big)\right]\\
& \Longrightarrow\quad \beta_p=(1-4\nu)p-\frac{p\,\beta_f}{\nu+f} \ .
\end{split}
\end{equation}}
\begin{equation}
    \begin{split}
        &\beta_f=f(f-2\nu) \ ,\\
        &\beta_p=(1-4\nu)p \ -\ \frac{p\,\beta_f}{\nu+f}\ =\ (1-4\nu)p+2fp+\mathcal{O}(pf^2) \ .
    \end{split}
\end{equation}
The fixed points of the theory then are
\begin{itemize}
    \item $f=p=0$,
    \item $f=2\nu, \ \ p=0$,
\end{itemize}
This shows that in this case we can still have DCFTs, in which, however, the coupling $p$ of the quartic deformation remains trivial. We show the resulting phase diagram of the theory in Fig. \ref{fig:phasediagram}.

\section{Fermionic $QED_4$ with a Yukawa interaction}
\label{sec:QED_Yukawa}
Having summarized the  framework and methodology in the previous section, we now turn our attention to fermionic $QED_4$ in the presence of a real, uncharged scalar coupled to the fermion via a Yukawa interaction.  The Lagrangian  is
\begin{equation}
    \label{eq:FermionicQED_Yukawa}
    \mathcal{L}=-\frac{1}{4e^2}F_{\mu\nu}F^{\mu\nu}+i\bar{\psi}\slashed{D}\psi+\frac{1}{2}\partial_{\mu}\phi\partial^{\mu}\phi+\frac{\lambda}{4}\phi
    ^4-y{\bar{\psi}\phi\psi}
\end{equation}
where $D_{\mu}=\partial_{\mu}-iA_{\mu}$, $\psi$ is a $4d$ Dirac fermion and $\phi$ a real scalar. Their masses are tuned to zero, so that the theory is classically conformal. By rescaling the fields as $\displaystyle{\psi=\frac{\Psi}{e} \ \text{and} \ \phi=\frac{\Phi}{e}}$ and adding a Wilson Line in the system, our action becomes
\begin{equation}
    \label{eq:FermionicQED_Yukawa_Rescaled}
    \mathcal{S}=\frac{1}{e^2}\int d^4 x\left\{-\frac{1}{4}F_{\mu\nu}F^{\mu\nu}+i\bar{\Psi}\slashed{D}\Psi+\frac{1}{2}\partial_{\mu}\Phi\partial^{\mu}\Phi+\frac{\lambda}{4e^2}\Phi^4-\frac{y}{e}\bar{\Psi}\Phi\Psi-e^2q\delta^3(\vec{x})\delta^{\mu}_0A_{\mu}(x)\right\}\ .
\end{equation}
As noted earlier, in the limit
\begin{equation}
    \label{eq:FermQED_Yukawa_Limit}
    \begin{split}
        &e\to0, \ \ \Tilde{\lambda}=\frac{\lambda}{e^2}=\text{fixed}, \ \ \Tilde{y}=\frac{y}{e}=\text{fixed}, \ \ Q=e^2q=\text{fixed}\\
        &\lambda\to0, \ \ y\to0, \ \ q\to\infty\ ,
    \end{split}
\end{equation}
we can work in the semiclassical approximation and the Landau pole of the $U(1)$ gauge coupling moves to infinity, so that we can ignore RG flows of the theory in the bulk. Here the theory remains conformal also at the quantum level. We can hence proceed by solving the EOMs
\begin{equation}
    \label{eq:FermQED_Yukawa_EOM}
    \begin{split}
        &\partial_{\mu}F^{\mu\nu}=-j^{\nu}+Q\delta^3(\vec{x})\delta^{\nu}_0,\\
        &(i\slashed{D}-\Tilde{y}\Phi)\Psi=0,\\
        &( \Box-\Tilde{\lambda}\Phi^2) \Phi =-\Tilde{y}\bar{\Psi}\Psi\ .
    \end{split}
\end{equation}
The leading order solution to these equations is
\begin{equation}
    \label{eq:SolFermQED_Yukawa}
    A_0(x)=\frac{Q}{4\pi r}, \ \ \Psi=0, \ \ \Phi=0.
\end{equation}
The quantum fluctuations around these field configurations  for $r\sim0$ are obtained by inserting the classical solution for the gauge field $A_{\mu}(x)$ into the EOMs \eqref{eq:FermQED_Yukawa_EOM}. Since the fermion field is charged under the $U(1)$ gauge symmetry, it is the first field to receive corrections. Via  the Yukawa operator $\Phi\bar{\Psi}\Psi$ the corrections to the fermion field will then impact  the scalar quantum fluctuations.\\
To determine these quantum fluctuations we will follow the steps in subsection \ref{subses:FermionicQED}. Exploiting the Weyl invariance of the theory we work directly in  $AdS_2\times S^2$. We recall, for reader's convenience, the expansion for the spinor field in $d$-dimensional space time 
\begin{equation}
\label{eq:ferm_exp_2}
    \Psi=\frac{1}{r^{\frac{d-1}{2}}}\sum_{\ell, s}\sum_{\delta=+,-}\psi^{(\delta)}_{\ell s}(t,r)\otimes \chi^{(\delta)}_{\ell s}(\hat{n}),
\end{equation}
where $\psi_{\ell s}^{(\delta)}(t,r)$ are $AdS_2$ spinors with 2 complex components and $\chi^{(\delta)}_{\ell s}(\hat{n})$ are the complete set of $2^{\lfloor\frac{d}{2} \rfloor-1}$-dimensional eigenstates of the Dirac equation on the $(d-2)$-dimensional sphere.    These quantities satisfy the conditions   \eqref{eq:spinor_harmonics} and \eqref{eq:harm_orth}. 

Before moving to the AdS description it is convenient to separate the action in four-dimensional flat space time \eqref{eq:FermionicQED_Yukawa_Rescaled} as
\begin{equation}
    \label{eq:action_separated}
    \begin{split}      
    &\mathcal{S}=\frac{1}{e^2}\Bigl[\mathcal{S}_{\text{bulk}}-e^2q\int dt A_0(x)\Bigr],\\
    &\mathcal{S}_{\text{Maxwell}}=-\frac{1}{4}\int d^4 x F^2,\\
    &\mathcal{S}_{\text{Ferm}}=i\int d^4 x \bar{\Psi}(\slashed{\partial}-i\slashed{A})\Psi\\
    &\mathcal{S}_{\text{Yuk}}=\int d^4 x \Big[\frac{1}{2}\partial_{\mu}\Phi\partial^{\mu}\Phi+\frac{\lambda}{4e^2}\Phi^4-\frac{y}{e}\bar{\Psi}\Phi\Psi\Bigr]
    \end{split}
\end{equation}
with $\mathcal{S}_{\text{bulk}}=\mathcal{S}_{\text{Maxwell}}+\mathcal{S}_{\text{Ferm}}+\mathcal{S}_{\text{Yuk}}$.
Following the detailed computations reported in the Appendix \ref{subsec:details_fermion}, the fermionic action takes the form
\begin{equation}
\label{eq:ferm_AdS2}
    \mathcal{S}_{\text{Ferm}} \to    \mathcal{S}_{\text{Ferm}}=\sum_{\ell s}\sum_{\delta=\pm}\int_{AdS_2}d^2x \sqrt{g}\bar{\psi}^{(\delta)}_{\ell s}[i(\slashed{\nabla}_{AdS_2}-i\slashed{A})-m_{\ell}]\psi^{(\delta)}_{\ell s}
\end{equation}
as in eq. \eqref{eq:final_ferm_S}. We note also that the field redefinition
\begin{equation}
\label{eq:psi_transf}
    \begin{split}
        &\psi^{(\delta)}_{\ell s}\to e^{-\delta i \frac{\pi}{4}\gamma^3} \psi^{(\delta)}_{\ell s}\\
        &\bar{\psi}^{(\delta)}_{\ell s}\to (\psi^{(\delta)}_{\ell s})^{\dagger}e^{\delta i \frac{\pi}{4}\gamma^3} \gamma^0
    \end{split}
\end{equation}
(where $\delta=\pm$) used to arrive to the form in eq. \eqref{eq:ferm_AdS2} further  modifies the scalar field EOM, since the fermion bilinear $\bar{\Psi}\Psi$ assumes the following form 
\begin{equation}
    \label{eq:ferm_bilin_transf}
    \bar{\Psi}\Psi\to\frac{1}{r^{d-1}}\sum_{\ell,\ell'}\sum_{s,s'}\left[\sum_{\delta\neq\delta'}(\psi^{(\delta)\dagger}_{\ell s}\sigma_1\psi^{(\delta')}_{\ell' s'})\chi^{(\delta)\dagger}_{\ell s}\chi^{(\delta')}_{\ell' s'}+\sum_{\delta=\delta'}\delta(\psi^{(\delta)\dagger}_{\ell s}\sigma_3\psi^{(\delta')}_{\ell' s'})\chi^{(\delta)\dagger}_{\ell s}\chi^{(\delta')}_{\ell' s'}\right] \ .
\end{equation}
The solution of the Dirac equation on $AdS_2$ for the fermionic quantum fluctuations, borrowed from the previous subsection \ref{subses:FermionicQED} (equations \eqref{eq:spinor_AdS} to \eqref{eq:nu_ell_fermion}) is

\begin{equation}
\label{eq:psi_sol_2}
    \psi^{(\delta)}_{\ell s}=\begin{pmatrix}
        \zeta^{(\delta)}_{\ell s} \\
        \xi^{(\delta)}_{\ell s}
    \end{pmatrix}
\end{equation}
with $\zeta$ and $\xi$ both single-component Grassmann fields, whose solution is
\begin{equation}
\label{eq:ferm_sol_yuk}
    \begin{split}
        &\zeta^{(\delta)}_{\ell s} = \alpha^{(\delta)}_{\ell s} r^{\frac{1}{2}-\nu_{\ell}} + \frac{g}{m_{\ell}+\nu_{\ell}}\beta^{(\delta)}_{\ell s}r^{\frac{1}{2}+\nu_{\ell}}\\
        &\xi^{(\delta)}_{\ell s} = \beta^{(\delta)}_{\ell s}r^{\frac{1}{2}+\nu_{\ell}}+\frac{g}{m_{\ell}+\nu_{\ell}}\alpha^{(\delta)}_{\ell s}r^{\frac{1}{2}-\nu_{\ell}}
    \end{split}
\end{equation}
with 
\begin{equation}
    \nu_{\ell}=\sqrt{m^2_{\ell}-g^2}
\end{equation}
in $d=4$. Having achieved a non-zero fermion field, we now determine the quantum fluctuations of the field $\Phi$ via its linearized EOM 
\begin{equation}
    \label{eq:scalar_eq_linearised}
    \Box \Phi = -\Tilde{y}\bar{\Psi}\Psi,
\end{equation}
using the $\bar{\Psi}\Psi$ bilinear as a source term. The reader will find in Appendix \ref{subsec:details_yukawa} the computational details needed to solve the previous equation, while we summarize here the salient steps. We first compute the products of spinor harmonics $\chi^{(\delta)\dagger}_{\ell s}\chi^{(\delta')}_{\ell' s'}$ in eq. \eqref{eq:ferm_bilin_transf}, by exploiting the fact that they are related to the canonical spherical spinors by a unitary transformation $V$ \cite{Abrikosov:2002jr} such that
\begin{equation}
    \label{eq:spin_harm_transf}
    \begin{split}
        V^{\dagger}\chi^{(\delta)}_{js}&=\frac{1}{\sqrt{2}}(\Omega_{j,j^-,s}-\delta \Omega_{j,j^+,s})=\\
        &=\frac{1}{\sqrt{2}}\begin{pmatrix}
         \sqrt{\frac{j+s}{2j}} Y_{j-\frac{1}{2},s-\frac{1}{2}} +\delta \sqrt{\frac{j-s+1}{2j+2}}Y_{j+\frac{1}{2},s-\frac{1}{2}}\\
          \sqrt{\frac{j-s}{2j}} Y_{j-\frac{1}{2},s+\frac{1}{2}} - \delta \sqrt{\frac{j+s+1}{2j+2}}Y_{j+\frac{1}{2},s+\frac{1}{2}}
        \end{pmatrix}, \\
        &j^{\pm}=j\pm \frac{1}{2}
    \end{split}
\end{equation}
where we re-labeled the spinors with their half-integer total angular momenta $j=\displaystyle{\ell+\frac{1}{2}}>0$ instead of the orbital one, $\ell\geq0$.\\
To solve eq. \eqref{eq:scalar_eq_linearised}, we decompose the scalar field in spherical harmonics as
\begin{equation}
    \Phi=\sum_{L,M}\phi_{L M}(r,t)Y_{L M}(\theta,\varphi).
\end{equation}
Inserting this expansion into the EOM for $\Phi$ and focusing on the $L=0$ mode,  total angular-momentum conservation forces the double sums in eq.~\eqref{eq:ferm_bilin_transf} to collapse to a single sum and the first term to become null, as explained in Appendix \ref{subsec:details_yukawa}. We hence find the EOM for $\phi_{L=0}$ to be\footnote{To avoid a cumbersome notation, from here we drop the $L, \ M$ indices in the  $J$ coefficients, as their value is now fixed to zero. We refer the reader to Appendix \ref{subsec:details_yukawa} for a further explanation of this equation.}
\begin{equation}
    \label{eq:zero_mode_EOM}
    \frac{2}{r}\phi'+\phi''=-\frac{\Tilde{y}}{\sqrt{4\pi}}\frac{1}{r^3}\sum_s\sum_{\ell}\sum_{\delta}J^{(\delta)}_{\ell  s}( \psi^{(\delta)\dagger}_{\ell  s}\sigma_3\psi^{(\delta)}_{\ell  s}) \ ,
\end{equation}
where we  neglected the time dependence of the scalar similarly to the spinor field. The prime over $\phi$ is a partial derivative with respect to $r$.  In Appendix \ref{subsec:details_yukawa} we report the values of the coefficients $J^{(\delta)}_{LM\ell_1\ell_2s_1s_2}$, stemming from manipulating the Clebsch-Gordan coefficients. We can finally write the solution of eq. \eqref{eq:zero_mode_EOM} as
\begin{equation}
\label{eq:final_sol}
    \phi_{L=0}(r)=\frac{\Tilde{y}}{\sqrt{4\pi}}\sum_s\sum_{\ell}\sum_{\delta}J^{(\delta)}_{\ell s}F^{(\delta)}_{\ell  s}(r)  \ \ \ \ (r\sim0) \ ,
\end{equation}
with 
\begin{equation}
\label{eq:F2_l}
\begin{split}
    F_{\ell}(r)&= \left[1-\frac{g^2}{(m_{\ell}+\nu_{\ell})^2}\right]\left[\alpha^{\dagger}_{\ell}\alpha_{\ell}\,\frac{r^{-2\nu_{\ell}}}{ 2\nu_{\ell}(1-2\nu_{\ell})}+\beta^{\dagger}_{\ell}\beta_{\ell}\,\frac{r^{2\nu_{\ell}}}{2\nu_{\ell}(1+2\nu_{\ell})}\right]\\
    &=\frac{1}{m_{\ell}+\nu_{\ell}}\left[\frac{\alpha^{\dagger}_{\ell}\alpha_{\ell}\,r^{-2\nu_{\ell}}}{1-2\nu_{\ell}}+\frac{\beta^{\dagger}_{\ell}\beta_{\ell}\,r^{2\nu_{\ell}}}{1+2\nu_{\ell}}\right] \ .
\end{split}
\end{equation}
Here we omitted the indices $\delta$ and $s$ to ease the notation. As shown in Appendix~\ref{subsec:details_yukawa}, for the $L=0$ mode the orthonormality of the spinor harmonics fixes $J^{(\delta)}_{\ell s}=\delta$ for all $\ell$ and $s$.  We again refer to Appendix \ref{subsec:details_yukawa} for the solutions with general $L$. 

\section{The modified RG flow with a Yukawa interaction: new fixed points}
\label{sec:new_RG}
\begin{figure}[h!]
\begin{center}
\includegraphics[width=9cm]{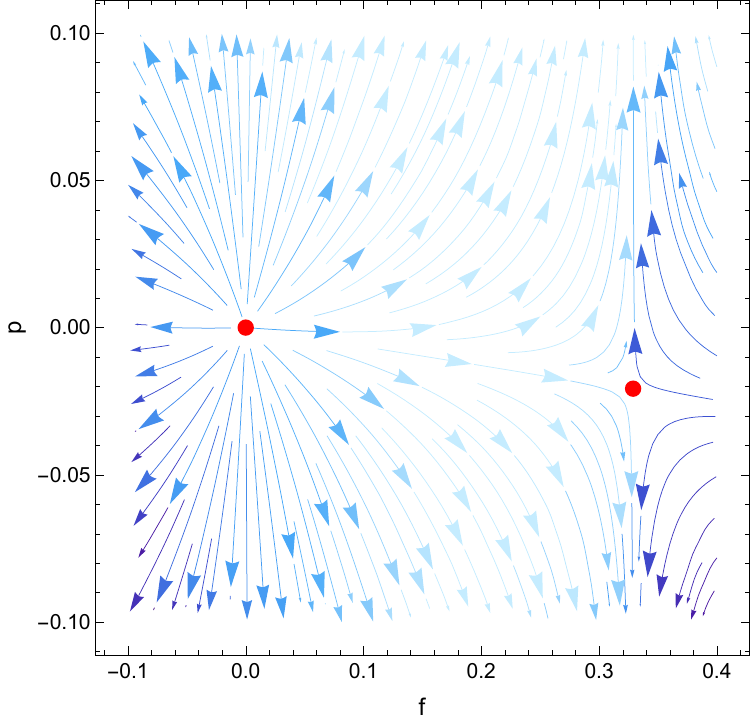}
\end{center}
\caption{\label{fig:phasediagram2}Phase diagram of Fermion QED$_4$ with Yukawa interaction, with quadratic and quartic defect deformations. Here we used $\nu=0.49$ and $\tilde{y}=0.5$. Arrows point towards the infrared: the origin is an infrared-repulsive node, while the interacting fixed point is a saddle, infrared-attractive along $f$ and repulsive along $p$ (eigenvalues $2\nu$ and $1-4\nu$ in $d/d\log\mu$). The plotted vector field is exactly $(\beta_f,\beta_p)$ of eq.~\eqref{eq:beta_p_corr} restricted to the $h=k=q=0$ slice, so the red dots are the zeros of the very field shown: the running-induced term $\Delta\beta_p^{\rm run}$ vanishes at the fixed points together with its $p$-derivative, whereas its expanded form $3(m+\nu)fp$ would incorrectly turn the saddle into a stable node. The $\beta_p$ fixed point $p_*$ is quoted in the reduced scheme; the exact-in-$F$ scheme rescales it by $\tfrac13$ (see Appendix~\ref{app:Yukawa_details}).}
\end{figure}

Following what was done for QED summarized in Section~\ref{sec:review} we now investigate the appropriate boundary terms to add to our starting bulk action in eq. \eqref{eq:action_separated} in $AdS_2\times S^2$. These terms will impact on the possible DCFTs we can have. Additionally we focus on the window $0<\nu_{\ell}<1/2$, where the fermionic solution in eq. \eqref{eq:ferm_sol_yuk} is normalizable.\\
The variation of our bulk action gives the boundary term
\begin{equation}
    \label{eq:bdy_variation}
    \delta S_{\text{bulk}}=\int_{r=r_0}dt\sqrt{-g}\left\{\frac{\delta \mathcal{L}}{\delta\partial_r\Phi}\delta\Phi+\frac{\delta \mathcal{L}}{\delta\partial_r\Psi}\delta\Psi+\delta\Bar{\Psi}\frac{\delta \mathcal{L}}{\delta\partial_r\Bar{\Psi}}\right\},
\end{equation}
which is evaluated for a small cut-off $r_0$. We also note that, for completeness, one should take into account the quadratic fluctuations of the gauge field $A_{\mu}$ and their contribution to $\delta S_{\text{bulk}}$. However, we checked that this contribution is actually not relevant for the analysis. We discuss this in more detail in Appendix \ref{app:A_m_deformations}.\\
In computing the variation in eq. \eqref{eq:bdy_variation}, we  focus on the $\ell=0$ modes for both the spinor and the scalar, as those are the ones carrying the leading powers in $r_0$. We, hence, separate eq. \eqref{eq:bdy_variation} into the two terms
\begin{equation}
    \delta S_{\text{bulk}}[\Psi]=-\frac{i}{2}\sum_{\delta=\pm}\sum_{s=\pm\frac{1}{2}}\int_{r_0}dt\sqrt{-g g^{rr}}(\psi^{(\delta)\dagger}_s\gamma^1\delta \psi^{(\delta)}_s-\delta\psi^{(\delta)\dagger}_s\gamma^1\psi^{(\delta)}_s),
\end{equation}
and
\begin{equation}
\label{eq:84}
    \delta S_{\text{bulk}}[\Phi]=\int_{r_0}dt\sqrt{-g}g^{rr}(\partial_r\Phi)\delta\Phi,
\end{equation}
where for the first one we borrow the results of subsection \ref{subses:FermionicQED} and write it as
\begin{equation}
   \delta S_{\text{bulk}}[\Psi] =\frac{\nu}{m+\nu}\sum_{\delta=\pm}\sum_{s=\pm\frac{1}{2}}\int_{r_0} dt(\beta_s^{(\delta)\dagger}\delta\alpha_s^{(\delta)}-\alpha^{(\delta)\dagger}_s\delta\beta_s^{(\delta)}+\delta\alpha_s^{(\delta)\dagger}\beta_s^{(\delta)}-\delta\beta_s^{(\delta)\dagger}\alpha_s^{(\delta)}).
\end{equation}
Regarding the scalar contribution, the leading contributions for $r\sim0$ are given by the terms with $\ell_1=\ell_2=0$, which come from the fermion modes with the lightest $AdS_2$ mass $m_{\ell}$. We are hence left with a bulk action variation
\begin{equation}
\label{eq:bulk_var}
\begin{split}
    \delta S_{\text{bulk}}=&\int_{r_0}dt\frac{\nu}{m+\nu}\sum_{\delta}\sum_s(\beta_s^{(\delta)\dagger}\delta\alpha_s^{(\delta)}-\alpha^{(\delta)\dagger}_s\delta\beta_s^{(\delta)}+\delta\alpha_s^{(\delta)\dagger}\beta_s^{(\delta)}-\delta\beta_s^{(\delta)\dagger}\alpha_s^{(\delta)})+\\
    &+\frac{\Tilde{y}^2}{4\pi(m+\nu)^2}\sum_{s_1,s_2}\sum_{\delta_1,\delta_2}J^{(\delta_1)}_{s_1}J^{(\delta_2)}_{s_2}\left\{\frac{r_0^{1-4\nu}}{1-2\nu}\alpha^{(\delta_1)\dagger}_{s_1}\alpha^{(\delta_1)}_{s_1}\delta(\alpha_{s_2}^{(\delta_2)\dagger}\alpha^{(\delta_2)}_{s_2})+\right.\\
    &+\frac{r_0}{1+2\nu} \alpha_{s_1}^{(\delta_1)\dagger}\alpha^{(\delta_1)}_{s_1}\delta(\beta_{s_2}^{(\delta_2)\dagger}\beta^{(\delta_2)}_{s_2})+\frac{r_0}{1-2\nu}\beta_{s_1}^{(\delta_1)\dagger}\beta^{(\delta_1)}_{s_1}\delta( \alpha_{s_2}^{(\delta_2)\dagger}\alpha_{s_2}^{(\delta_2)})+\\
    &+\left.\frac{r_0^{1+4\nu}}{1+2\nu}\beta_{s_1}^{(\delta_1)\dagger}\beta^{(\delta_1)}_{s_1}\delta(\beta^{(\delta_2)\dagger}_{s_2}\beta_{s_2}^{(\delta_2)}) \right\}+\text{sub.}
\end{split}
\end{equation}
Defining $\displaystyle{\alpha\coloneqq\big(\alpha^{(+)}_{\frac{1}{2}} \ \alpha^{(+)}_{-\frac{1}{2}} \ \alpha^{(-)}_{\frac{1}{2}} \ \alpha^{(-)}_{-\frac{1}{2}}\big)^T}$ (and $\beta$ analogously), we can group the terms as

\begin{equation}
\label{eq:bulk_var2}
    \begin{split}
        \delta \mathcal{S}_{\text{bulk}}
&=
\int dt\,
\Biggl[
\frac{\nu}{m+\nu}
\left(
\bar{\beta}\,\delta\alpha
-
\bar{\alpha}\,\delta\beta
+\text{h.c.}
\right)\\
&+\,
\frac{\tilde{y}^{\,2}}{4\pi(m+\nu)^{2}}
\Biggl[
\frac{r_0^{\,1-4\nu}}{1-2\nu}
\Big((\bar{\alpha}J\alpha)(\bar{\alpha}J\delta\alpha)
+
(\bar{\alpha}J\alpha)
(\delta\bar{\alpha}J\alpha)\Big)\\
&+\,
\frac{r_0}{1+2\nu}
\,
\bar{\alpha}J\alpha\,
\delta(\bar{\beta}J\beta)
+
\frac{r_0}{1-2\nu}
\,
\delta(\bar{\alpha}J\alpha)
(\bar{\beta}J\beta)
+\text{sub.}
\Biggr]
\Biggr] \ .
    \end{split}
\end{equation}
with $J
\coloneqq
\operatorname{diag}
\left(
J_{s_1}^{(\delta_1)},
J_{s_2}^{(\delta_1)},
J_{s_1}^{(\delta_2)},
J_{s_2}^{(\delta_2)}
\right)  = \operatorname{diag}(1,1,-1,-1)=\Sigma^{q_0}$, as derived in Appendix~\ref{subsec:details_yukawa}, eq. \eqref{eq:J_corrected}.  In the range $\displaystyle{0<\nu<\frac{1}{2}}$  the field configurations for the boson and for the spinors are normalisable. It is useful to further split this range to highlight, as we shall see, new interesting dynamics occurring due to the presence of the neutral scalar field. In fact, in the range $\displaystyle{0<\nu<\frac{1}{4}}$, the leading terms are the first four present in eq. \eqref{eq:bulk_var} and the only relevant operator is $\alpha^{\dagger}\alpha$ ($\Delta(\alpha^{\dagger}\alpha)=1-2\nu$). 
 However, for $\displaystyle{\frac{1}{4}<\nu<\frac{1}{2}}$, the quartic operators\footnote{Because of the fermionic statistics, however, only a subset of these operators will be non-zero (i.e. the ones in which there are no repeating modes). In this work we will limit our analysis to 4-fermion operators, but it is also worth mentioning that one could carry the analysis further and consider 8-fermion operators, which become relevant for $\nu>3/8$, or $\frac{e^2q}{4\pi}<\frac{\sqrt{55}}{8}$. Because of Fermi statistics, however, there will be no terms containing more than 8 fermion modes, as each of them carries two indices ($\delta$ and $s$) which would repeat in such a term, setting it to zero.} $\alpha^{(\delta)\dagger}_s\alpha^{(\delta)}_s\alpha^{(\delta')\dagger}_{s'}\alpha^{(\delta')}_{s'}$ (and similar) become relevant making it interesting to study the effect of multi-trace deformations on the bulk action. This is the regime we will be studying in what follows.\\
From eq. \eqref{eq:bulk_var2} we obtain the boundary condition
\begin{equation}
  (\bar \alpha J \alpha)J\alpha=0   \ .
\end{equation}
We can modify this condition via the boundary term
\begin{equation}
    \mathcal{S}_{\text{bdy}}
=
\mathcal{S}_{\text{bdy}}^{(1)}
-
\frac12
\int dt\,
\frac{\tilde{y}^{\,2}\,
r_0^{\,1-4\nu}}
{4\pi(m+\nu)^{2}(1-2\nu)}
(\bar{\alpha}J\alpha)^2 \ ,
\end{equation}
where $\mathcal{S}_{\text{bdy}}^{(1)}$ the one in eq. \eqref{eq:ferm_bdy_1}. With this, the boundary condition becomes
\begin{equation}
    \delta
\left(
\mathcal{S}_{\text{bulk}}
+
\mathcal{S}_{\text{bdy}}
\right)
=
0
\qquad\Longleftrightarrow\qquad
\beta=\bar{\beta}=0 \ .
\end{equation}
As for scalar and fermionic QED$_4$, this boundary condition means that the theory contains relevant operators that, as mentioned above, are of the form $\alpha^2$ and $\alpha^4$. We can study the RG flows triggered by these operators by adding the double trace deformation $\mathcal{S}_{\mathrm{DTD}}$ in eq. \eqref{eq:bdy_fermion_4d} and the quartic deformation\footnote{For completeness, let us observe that this is only one of the possible choices we can make. The most generic quartic deformation is generated from building blocks of the sort $(\bar{\alpha} \Sigma^A \alpha)$ (the $\Sigma^A$'s being the matrices in eq. \eqref{eq:sigmas}). However, in the following we only use $\Sigma^A=\mathds{1}$ to keep the discussion simple.}
\begin{equation}
    \mathcal{S}_{MTD} \;=\; -p_0 \, r_0^{4\nu-1} \int dt \,
\Bigg\{
\big(\bar{\alpha}\,\alpha\big)\big(\bar\alpha\,\alpha\big)\, r_0^{1-4\nu}
+\; \big(\bar\alpha\,\alpha\big)\big(\bar\alpha\beta + \bar\beta\alpha\big)\, r_0^{1-2\nu}
\Bigg\} \ 
\end{equation}
such that $\displaystyle{\mathcal{S}_{MTD}\to -p_0\int dt \alpha^4}$ for $r_0\to 0$.\\
Now, the condition $\delta\big(\mathcal{S}_{\text{bulk}} + \mathcal{S}_{\text{bdy}} + \mathcal{S}_{DTD} + \mathcal{S}_{MTD}\big) = 0$ implies
\begin{equation}
\label{eq:cond_a}
    \begin{split}
        \frac{2\nu}{m+\nu}\,\beta \;-\; 2 r_0^{2\nu} F_0 \beta \;-\; 2 F_0 \alpha
\;-\; 2 p_0 \alpha\,\bar\alpha\alpha \;-\; p_0 r_0^{2\nu}\,\alpha\bar\beta\alpha\\
-\; p_0 r_0^{2\nu}\,\alpha\bar\alpha\beta \;-\; p_0 r_0^{2\nu}\,\beta\bar\alpha\alpha
\;+\; \frac{\tilde{y}^2 r_0}{4\pi(m+\nu)^2(1-2\nu)}\, J\alpha\,\big(\bar\beta J \beta\big) \;=\; 0
    \end{split}
\end{equation}
and
\begin{equation}
\label{eq:cond_b}
    \begin{split}
        \frac{2\nu}{m+\nu}\,\bar\beta \;-\; 2 r_0^{2\nu}\,\bar\beta F_0 \;-\; 2\bar\alpha F_0
\;-\; 2 p_0\,\bar\alpha\bar\alpha\alpha \;-\; r_0^{2\nu} p_0\,\bar\alpha\bar\alpha\beta\\
-\; p_0 r_0^{2\nu}\,\alpha\bar\beta\alpha \;-\; p_0 r_0^{2\nu}\,\bar\beta\,\bar\alpha\alpha
\;+\; \frac{\tilde{y}^2 r_0}{4\pi(m+\nu)^2(1-2\nu)}\,(\bar\alpha J)\big(\bar\beta J\beta\big) \;=\; 0
    \end{split}
\end{equation}
From these, keeping only linear terms in $p$ and quadratic ones in $F$, we get the boundary condition\footnote{The interested reader can find the details of the computation in Appendix \ref{app:Yukawa_details}.}
\begin{equation}
    \begin{split}
       \beta &= \frac{(m+\nu) r_0^{-2\nu}}{\nu}
\left(\mathds{1} + \frac{m+\nu}{\nu} F\right) F\alpha
\;+\; \frac{m+\nu}{\nu}\, p\, r_0^{1-4\nu}\,\alpha(\bar\alpha\alpha)
\;+\; \frac{(m+\nu)^2}{\nu^2}\, r_0^{-4\nu}\, p\,\alpha(\bar\alpha F\alpha) \;+ \\
&+\; \frac{(m+\nu)^2}{2\nu^2}\, r_0^{1-4\nu}\, p\,(F\alpha)(\bar\alpha\alpha)
\;-\;\frac{\tilde{y}^2\, r_0^{1-4\nu}}{2\pi\nu(m+\nu)(1-2\nu)}\,(J\alpha)(\bar\alpha F J F\alpha) \;+\\
&+\; \frac{(m+\nu)^3}{2\nu^3}\, p\, r_0^{1-4\nu}\,(F\alpha)(\bar\alpha F\alpha)+\text{sub.}
    \end{split}
\end{equation}
Imposing that this does not depend on the renormalization scale $\mu=r_0^{-1}$, we get the beta functions
\begin{equation}
   \beta_F=-2\nu F+2(m+\nu)F\cdot F \ ,
\end{equation}
and
\begin{equation}
\label{eq:beta_p_corr}
    \beta_p = (1-4\nu)p\;+\;\Delta\beta_p^{\rm run}\;+\; \frac{(4 \nu -1)\, \tilde y^2 \left(f^2-h^2-k^2+q^2\right)}{6 \pi (m+\nu)^2 (2 \nu -1)}\ ,
\end{equation}
with $p=r_0^{4\nu-1}p_0$ and
\begin{equation}
\label{eq:delta_beta_run}
 \Delta\beta_p^{\rm run}\;=\;-\,p\,\frac{m+\nu}{12\nu}\Big[\operatorname{Tr}\beta_F\Big(\operatorname{Tr}B+\tfrac12\operatorname{Tr}A\Big)-\operatorname{Tr}(\beta_F B)-\tfrac12\operatorname{Tr}(A\beta_F)\Big]\ ,
\end{equation}
where $A=\mathds{1}+\frac{m+\nu}{\nu}F$ and $B=\mathds{1}+\frac{m+\nu}{2\nu}F$ (see Appendix~\ref{app:Yukawa_details}). To find the fixed points of the theory we notice that $\beta_F$ is completely independent of $p$ and has the set of fixed points discussed in Section \ref{subses:FermionicQED}. By substituting them in $\beta_p$, we find the set of fixed points of the theory:
\begin{itemize}
    \item $(f,k,h,q,p)=(0,0,0,0,0)$;
    \item $(f,k,h,q,p)=\Big(\frac{\nu}{m+\nu},0,0,0,\frac{\nu^2 \tilde y^2}{6 \pi  (2 \nu -1) (m+\nu )^4}\Big)$;
    \item $f=\frac{\nu}{2(m+\nu)}$, $k^2+h^2+q^2=\frac{\nu^2}{4(m+\nu)^2}$, $p=\frac{q^2 \tilde y^2}{3\pi(2\nu-1)(m+\nu)^2}$. On this family one has $f^2-h^2-k^2+q^2=2q^2$, so that the quartic coupling vanishes on the $q=0$ circle.
\end{itemize}
As discussed in Appendix \ref{app:Yukawa_details}, the values reported here for the $\beta_p$ hold in a quadratic approximation in F. In the exact-in-$F$ scheme, the same trace equation gives instead $p=\frac{\nu^2\tilde y^2}{18\pi(2\nu-1)(m+\nu)^4}$, one third of the reduced value ($\simeq-0.0108$ versus $-0.0323$ at the parameters of Fig.~\ref{fig:phasediagram2}): this order-one scheme dependence is unavoidable for fixed points of $\beta_F$ $f_*=\mathcal{O}(1)$ and must accompany any quoted value of $p_*$.
In Fig. \ref{fig:phasediagram2} we show the phase diagram of the theory in the case $h=k=0$ and only $f$ and $p$ are present as defect deformations couplings.\\
In the scenario contemplated here where $\nu$ is in the range $\displaystyle{\frac{1}{4}<\nu<\frac{1}{2}}$ QED with Yukawa interactions still features fixed point dynamics in presence of a line defect. This is the interesting range of $\nu$ since, here, the Yukawa interactions are non-negligible in contrast with the $\displaystyle{0<\nu<\frac{1}{4}}$ case where the Yukawa sector can be neglected. The interesting effects occur via four-fermion operators. For $\nu=1/4$ the quartic operator $(\alpha^{\dagger}\alpha)^2$ is marginal while $\alpha^{\dagger}\alpha$ stays relevant. In such a case we can always remove the $(\alpha^{\dagger}\alpha)\delta(\alpha^{\dagger}\alpha)$ term from eq. \eqref{eq:bulk_var2} and additionally use the same boundary terms as the $\nu<1/4$ case to recover the running of Section \ref{subses:FermionicQED}.

\section{Conclusions}
In this work we first reviewed the defect induced conformal physics of QED$_4$ in the presence of an infinitely heavy state of charge $q$. We then moved to investigate the dynamics in the presence of Yukawa interactions spurred by the addition of a real scalar field. We observed the emergence of relevant four-fermion operators which trigger an RG flow on the Wilson line.  This occurs for a specific range of  $e^2 q$ once $y/e$ is fixed. We discover a new structure of unstable fixed points going beyond the case of ordinary QED$_4$. Beyond four-fermion operators the Fermi statistics only allows the possible effect of eight-fermion operators in the smaller range  $3/8<\nu<1/2$ which will be analyzed elsewhere. Future directions for this work include applying the framework to the study of interacting field theories, such as  asymptotically safe theories \cite{Litim:2014uca}. We added a series of appendices in which we report the details of our computations.

\paragraph{Acknowledgements}

\noindent
The work of G.M. and F.S. are partially supported by the Carlsberg Foundation, grant CF22-0922. The work of A.D.A. is funded by the Department of Physics “Ettore Pancini” under the project PON MAGIC - CUP B69J23000560005. 
\newpage
\begin{appendices}

\section{Details on Fermionic $QED_4$ with Yukawa interaction}
\label{sec:Yukawa_details}
\subsection{Fermionic $QED_4$}
\label{subsec:details_fermion}
In this section we will expand on the computations described in section \ref{subses:FermionicQED}. In particular, we will focus on the steps that lead to eq.  \eqref{eq:final_ferm_S}. In general space-time dimension $d$, we start from the decomposition 
\begin{equation}
\label{eq:fermion_exp_d}
    \Psi=\frac{1}{r^{\frac{d-1}{2}}}\sum_{\ell, s}\sum_{\delta=+,-}\psi^{(\delta)}_{\ell s}(t,r)\otimes \chi^{(\delta)}_{\ell s}(\hat{n}),
\end{equation}
where $\psi_{\ell s}^{(\delta)}(t,r)$ are $AdS_2$ spinors with $2$ complex components and $\chi^{(\delta)}_{\ell s}(\hat{n})$ are the $2^{\lfloor\frac{d}{2} \rfloor-1}$-dimensional spinor harmonics which satisfy the Dirac equation on the $(d-2)$-dimensional sphere $S^{d-2}$ \cite{Lopez-Ortega:2009flo, Camporesi:1995fb}
\begin{equation}
    \label{eq:spinor_harmonics}
    \slashed{\nabla}_{S^{d-2}}\chi^{(\delta)}_{\ell s}(\hat{n})=\delta i \left(\ell +\frac{d-2}{2}\right)\chi^{(\delta)}_{\ell s}(\hat{n}), \ \ \ \ \ \delta=\pm
\end{equation}
and the orthogonality relation
\begin{equation}
    \label{eq:harm_orth}
    \int_{S^{d-2}}d\Omega_{d-2} \chi^{(\delta)\dagger}_{\ell s}(\hat{n})\chi^{(\delta')}_{\ell' s'}(\hat{n})=\delta_{\ell \ell'}\delta_{s s'}\delta^{\delta \delta'}
\end{equation}
Moreover, in eq. \eqref{eq:Fermion_expansion} the sum in $\ell$ runs over integer angular momenta $\ell=0,1,\dots$, while the one in $s$ runs over the spinor harmonics components.\\
Performing the Weyl transformation in dimension $d=4$, we can hence map the theory from Minkowski space $\mathbb{R}^{1,3}$ to $AdS_2\times S^{2}$ space and using equations \eqref{eq:fermion_exp_d}-\eqref{eq:harm_orth} the Dirac action becomes 
\begin{equation}
\label{eq:fermion_Weyl_rescaling}
\begin{split}
    &\mathcal{S}=\int d^d x i\bar{\Psi}(\slashed{\partial}-i\slashed{A})\Psi \ \to \ \\
   & \ \to \ \sum_{\ell,\ell'}\sum_{s,s'}\sum_{\delta,\delta'}\int_{AdS_2} d^2x\int_{S^{2}}d\Omega_{2}i\sqrt{g}\psi^{(\delta)\dagger}_{\ell s} \otimes \chi^{(\delta)\dagger}_{\ell s}\Gamma^0(\slashed{\nabla}_{AdS^2}+\slashed{\nabla}_{S^{d-2}}-i\slashed{A})\psi^{(\delta')}_{\ell' s'} \otimes \chi^{(\delta')}_{\ell' s'}
\end{split}
\end{equation}
Using the gamma matrices
\begin{equation}
    \begin{split}
        &\Gamma^0=\gamma^0\otimes\mathds{1},\\
        &\Gamma^1=\gamma^1\otimes\mathds{1},\\
        &\Gamma^2=i\gamma^3\otimes\hat{\gamma}^1,\\
        &\Gamma^3=i\gamma^3\otimes\hat{\gamma}^2
    \end{split}
\end{equation}
chosen in such a way to satisfy a Clifford Algebra\footnote{The $2\times2$ gamma matrices $\gamma^{0,1}$ and $\hat{\gamma}^{1,2}$ are respectively a representation of the Clifford Algebra with Lorentzian signature, i.e. $\{\gamma^{a},\gamma^b\}=2\text{diag}(1,-1)$, and a representation of the Clifford Algebra with Euclidean signature, i.e. $\{\hat{\gamma}^i,\hat{\gamma}^j\}=2\delta^{ij}$.} $\{\Gamma^{\mu},\Gamma^{\nu}\}=2\eta^{\mu\nu}$, where $\gamma^3=\gamma^0\gamma^1$, and using the spinor harmonics orthogonality relation \eqref{eq:harm_orth} we get
\begin{equation}
  \label{eq:final_fermion}  
  \mathcal{S}= \sum_{\ell,s}\sum_{\delta=\pm}\int dt dr \sqrt{g}i\psi^{(\delta)\dagger}_{\ell s} \gamma^0 (\slashed{\nabla}_{AdS_2}-i\slashed{A}-\delta\gamma^3 m_{\ell})\psi^{(\delta)}_{\ell s} 
\end{equation}
with $m_{\ell}=\displaystyle{\ell+\frac{d-2}{2}=\ell+1}$ for $d=4$ and with $\displaystyle{\sqrt{g}=\frac{1}{r^2}}$, as the $AdS_2$ metric is
\begin{equation}
    g_{\mu\nu}=\begin{pmatrix}
        \frac{1}{r^2} & 0\\
        0 & -\frac{1}{r^2}
    \end{pmatrix}
\end{equation}
Eq. \eqref{eq:final_fermion} is the action from which we can derive the $AdS_2$ spinors. Before doing that, though, it is possible to bring this action to a more symmetric form by performing the field redefinition
\begin{equation}
\label{eq:ferm_trans}
    \begin{split}
        &\psi^{(\delta)}_{\ell s}\to e^{-\delta i \frac{\pi}{4}\gamma^3} \psi^{(\delta)}_{\ell s}\\
        &\bar{\psi}^{(\delta)}_{\ell s}\to (\psi^{(\delta)}_{\ell s})^{\dagger}e^{\delta i \frac{\pi}{4}\gamma^3} \gamma^0
    \end{split}
\end{equation}
where $\delta=\pm$. By doing this, we find, as promised, eq. \eqref{eq:final_ferm_S}.\\

\subsubsection{The Dirac equation on $AdS_2$}
\label{app:AdSS2_fermions}
In this section we briefly illustrate the steps that lead to eq. \eqref{eq:psi_eom}. The Dirac equation for a spinor coupled to a $U(1)$ background gauge field on a curved space is given by
\begin{equation}
    \label{eq:dirac_eq_curved}
    i\gamma^a e_a^{ \ \ \mu}(\nabla_{\mu}-iA_{\mu})\psi=0,
\end{equation}
where $e_a^{ \ \ \mu}$ are the vielbeins \cite{Parker_Toms_2009}, $\displaystyle{\nabla_{\mu}=\partial_{\mu}+\frac{1}{8}\omega_{\mu ab}[\gamma^a,\gamma^b]}$ is the spinor covariant derivative, and $\omega_{\mu ab}$ the spin connection. We are interested in the form this equation takes on $AdS_2$ space, with coordinates $\{t,r\}$ in which the metric is
\begin{equation}
    \label{eq:AdS2_metric}
    ds^2=\frac{dt^2-dr^2}{r^2} \ \ \Rightarrow \ \ g_{\mu\nu}=\begin{pmatrix}
        \frac{1}{r^2} & 0 \\
        0 & -\frac{1}{r^2}
    \end{pmatrix}
\end{equation}
Recalling that the vielbeins must satisfy the equation $g_{\mu\nu}=e^a_{ \ \ \mu}e^b_{ \ \ \nu}\eta_{ab}$, with $\eta_{ab}=\text{diag}(1,-1)$ a flat metric, we can choose them to be diagonal, and in particular
\begin{equation}
    e^a_{ \ \ \mu}=\begin{pmatrix}
        \frac{1}{r} & 0 \\
        0 & \frac{1}{r}
    \end{pmatrix}.
\end{equation}
From here we can proceed with the computation of the spin connection $\omega_{\mu}^{ \ \ ab}=-e_b^{ \ \ \nu}(\partial_{\mu}e^a_{ \ \ \nu}-\Gamma^{\lambda}_{\mu\nu}e^a_{ \ \ \lambda})$, with $\Gamma^{\lambda}_{\mu\nu}$ the Christoffel symbols. The only non-zero Christoffel symbols we have for the metric in eq. \eqref{eq:AdS2_metric} are
\begin{equation}
    \label{eq:AdS2_Chris_Symb}
    \Gamma^{1}_{00}=-\frac{1}{r}, \ \ \Gamma^{1}_{11}=-\frac{1}{r}, \  \ \Gamma^0_{01}=\Gamma^0_{10}=-\frac{1}{r}.
\end{equation}
We thus obtain that the only non-zero term of the spin connection is 
\begin{equation}
    \label{eq:AdS2_spin_conn}
    \omega_{0 \   1}^{  \ 1}=-\frac{1}{r}.
\end{equation}
Choosing a basis in which the Gamma matrices are 
\begin{equation}
    \label{eq:gamma_mat_AdS2}
    \gamma^0=\sigma_1, \ \ \gamma^1=i\sigma_3, 
\end{equation}
with $\sigma_i$ the Pauli matrices, we have finally all the ingredients to proceed in our computation. With the above, neglecting time dependence, we obtain
\begin{equation}
    \label{eq:cov_der_spin_AdS2}
    \slashed{\nabla}\psi=\Bigl[r\gamma^1\partial_r-\frac{1}{2}\gamma^1\Bigr]\psi.
\end{equation}
Hence, eq. \eqref{eq:dirac_eq_curved} becomes
\begin{equation}
    i\left(r\gamma^1\partial_r-\frac{1}{2}\gamma^1-ig\gamma^0\right) \psi(r)=0,
\end{equation}
which is eq. \eqref{eq:psi_eom} (minus the mass term), as promised.\\

\subsection{Adding the Yukawa interaction}
\label{subsec:details_yukawa}
In this section we will expand on the steps described in section \ref{sec:QED_Yukawa}, when computing the solution of the EOM for the scalar field $\Phi$, i.e. 
\begin{equation}
\label{eq:app__EOM_scalar}
    \Box \Phi = -\Tilde{y}\bar{\Psi}\Psi
\end{equation}
Let us hence commence by computing the bilinear in eq. \eqref{eq:ferm_bilin_transf} with the solution for $\psi^{(\delta)}_{\ell s}$ in eq. \eqref{eq:ferm_sol_yuk}. We have\footnote{We momentarily suppress the $\delta$ superscript and $s$ subscript to avoid cumbersome notation.} 
\begin{equation}
    \label{eq:psi_bar_psi}
    \begin{split}
        \psi^{\dagger}_{\ell_1}\sigma_1\psi_{\ell_2 }=&\alpha^{\dagger}_{\ell_1}\beta_{\ell_2}r^{1-\nu_{\ell_1}+\nu_{\ell_2}}+\frac{g}{m_{\ell_2}+\nu_{\ell_2}}\alpha^{\dagger}_{\ell_1}\alpha_{\ell_2}r^{1-\nu_{\ell_1}-\nu_{\ell_2}}+\frac{g}{m_{\ell_1}+\nu_{\ell_1}}\beta^{\dagger}_{\ell_1}\beta_{\ell_2}r^{1+\nu_{\ell_1}+\nu_{\ell_2}}+\\
        &+\frac{g^2}{(m_{\ell_1}+\nu_{\ell_1})(m_{\ell_2}+\nu_{\ell_2})}\beta^{\dagger}_{\ell_1}\alpha_{\ell_2}r^{1+\nu_{\ell_1}-\nu_{\ell_2}}+\beta^{\dagger}_{\ell_1}\alpha_{\ell_2}r^{1+\nu_{\ell_1}-\nu_{\ell_2}}+\\
        &+\frac{g}{m_{\ell_2}+\nu_{\ell_2}}\beta^{\dagger}_{\ell_1}\beta_{\ell_2}r^{1+\nu_{\ell_1}+\nu_{\ell_2}}+\frac{g}{m_{\ell_1}+\nu_{\ell_1}}\alpha^{\dagger}_{\ell_1}\alpha_{\ell_2}r^{1-\nu_{\ell_1}-\nu_{\ell_2}}+\\
        &+\frac{g^2}{(m_{\ell_1}+\nu_{\ell_1})(m_{\ell_2}+\nu_{\ell_2})}\alpha^{\dagger}_{\ell_1}\beta_{\ell_2}r^{1-\nu_{\ell_1}+\nu_{\ell_2}}
    \end{split}
\end{equation}
and 
\begin{equation}
    \label{eq:psi_bar_psi2}
    \begin{split}
         \psi^{\dagger}_{\ell_1}\sigma_3\psi_{\ell_2 }
        &=\alpha^{\dagger}_{\ell_1}\alpha_{\ell_2}r^{1-\nu_{\ell_1}-\nu_{\ell_2}}+\frac{g}{m_{\ell_2}+\nu_{\ell_2}}\alpha^{\dagger}_{\ell_1}\beta_{\ell_2}r^{1-\nu_{\ell_1}+\nu_{\ell_2}}+\frac{g}{m_{\ell_1}+\nu_{\ell_1}}\beta^{\dagger}_{\ell_1}\alpha_{\ell_2}r^{1+\nu_{\ell_1}-\nu_{\ell_2}}+\\
        &+\frac{g^2}{(m_{\ell_1}+\nu_{\ell_1})(m_{\ell_2}+\nu_{\ell_2})}\beta^{\dagger}_{\ell_1}\beta_{\ell_2}r^{1+\nu_{\ell_1}+\nu_{\ell_2}}-\beta^{\dagger}_{\ell_1}\beta_{\ell_2}r^{1+\nu_{\ell_1}+\nu_{\ell_2}}-\frac{g}{m_{\ell_2}+\nu_{\ell_2}}\beta^{\dagger}_{\ell_1}\alpha_{\ell_2}r^{1+\nu_{\ell_1}-\nu_{\ell_2}}+\\
        &-\frac{g}{m_{\ell_1}+\nu_{\ell_1}}\alpha^{\dagger}_{\ell_1}\beta_{\ell_2}r^{1-\nu_{\ell_1}+\nu_{\ell_2}}-\frac{g^2}{(m_{\ell_1}+\nu_{\ell_1})(m_{\ell_2}+\nu_{\ell_2})}\alpha^{\dagger}_{\ell_1}\alpha_{\ell_2}r^{1-\nu_{\ell_1}-\nu_{\ell_2}}
    \end{split}
\end{equation}
With this being said, we solve eq. \eqref{eq:app__EOM_scalar} by decomposing the scalar field as
\begin{equation}
\label{eq:phi_exp_2}
    \Phi=\sum_{L,M}\phi_{LM}(r,t)Y_{LM}(\theta,\varphi).
\end{equation}
Plugging this expansion into the EOM and using the relation 
\begin{equation}
\label{eq:harmonics_exp}
    Y_{\ell_1,m_1}Y_{\ell_2,m_2}=\sum_{L,M}\sqrt{\frac{(2\ell_1+1)(2\ell_2+1)}{4\pi(2L+1)}}c(\ell_1,\ell_2,0,0;L,0)c(\ell_1,\ell_2,m_1,m_2;L,M)Y_{L,M}
\end{equation}
where $c(j_1,j_2,m_1,m_2;J,M)=\braket{j_1,m_1;j_2,m_2|J,M}$ are the Clebsch-Gordan coefficients, the EOM for $\Phi$ becomes
\begin{equation}
\label{eq:EOM_phi_separating2}
\begin{split}
  &\sum_{L,M}\left(-\frac{L(L+1)}{r^2}+  \partial^2_r+\frac{2}{r}\partial_r\right)\phi_{LM}(r)Y_{LM}=\\
  &=-\frac{\Tilde{y}}{r^3}\sum_{LM}\frac{1}{\sqrt{4\pi(2L+1)}}Y_{LM}\sum_{\ell_1,\ell_2}\sum_{s_1,s_2}\Bigl[\sum_{\delta_1\neq\delta_2}K^{(\delta_1,\delta_2)}_{LM\ell_1\ell_2s_1s_2}( \psi^{(\delta_1)\dagger}_{\ell_1 s_1}\sigma_1\psi^{(\delta_2)}_{\ell_2 })+\sum_{\delta}J^{(\delta)}_{LM\ell_1\ell_2s_1s_2}( \psi^{(\delta)\dagger}_{\ell_1 s_1}\sigma_3\psi^{(\delta)}_{\ell_2 s_2})\Bigr]
\end{split}
\end{equation}
where we have neglected time dependence as for the spinor field solution.\\
In eq. \eqref{eq:EOM_phi_separating2} the values of the coefficients $K^{(\delta_1,\delta_2)}_{LM\ell_1\ell_2s_1s_2}$ and $J^{(\delta)}_{LM\ell_1\ell_2s_1s_2}$ are combinations of Clebsch--Gordan coefficients and read as follows. Calling $j_i=\ell_i+1/2>0, \ \text{with} \  \ell_i\geq0$ integers and $s_i=\pm 1/2, \ \dots$,
\begin{equation}
    \label{eq:K_coeff}
    \begin{split}
    K^{(\delta,\delta')}_{LM jj'ss'}&=\sqrt{4\pi(2L+1)}\int d\Omega \  Y^*_{LM} \chi^{(\delta) \dagger}_{\ell s} \chi^{(\delta')}_{\ell' s'}
    \end{split}
\end{equation}
and
\begin{equation}
    \label{eq:J_coeff}
    \begin{split}
    J^{(\delta)}_{LM jj'ss'}&=\delta\sqrt{4\pi(2L+1)}\int d\Omega \  Y^*_{LM} \chi^{(\delta) \dagger}_{\ell s} \chi^{(\delta)}_{\ell' s'} \ .
    \end{split}
\end{equation}
As will be discussed below, since our focus here will be on the $L=0$ mode of the scalar field, we only need the $\chi^{(\delta)\dagger}_{\ell s}\chi^{(\delta)}_{\ell s}$ product, which reads as follows (using eq. \eqref{eq:spin_harm_transf}):
\begin{equation}
    \label{eq:coeff_terms}
    \begin{split}
\chi^{(\delta)\dagger}_{\ell s}\chi^{(\delta)}_{\ell s} &= \frac{1}{2}\Bigg\{
\frac{j+s}{2j}(-1)^{s-\frac12} Y_{j-\frac12,-s+\frac12}\, Y_{j-\frac12,s-\frac12}
+ \frac{j-s+1}{2j+2}(-1)^{s-\frac12} Y_{j+\frac12,-s+\frac12}\, Y_{j+\frac12,s-\frac12}\\
&+ \delta\sqrt{\frac{j+s}{2j}}\sqrt{\frac{j-s+1}{2j+2}}\,(-1)^{s-\frac12}\, Y_{j-\frac12,-s+\frac12}\, Y_{j+\frac12,s-\frac12}\\
&+ \delta\sqrt{\frac{j+s}{2j}}\sqrt{\frac{j-s+1}{2j+2}}\,(-1)^{s-\frac12}\, Y_{j+\frac12,-s+\frac12}\, Y_{j-\frac12,s-\frac12} \\
&\quad + \frac{j-s}{2j}(-1)^{s+\frac12} Y_{j-\frac12,-s-\frac12}\, Y_{j-\frac12,s+\frac12}
+ \frac{j+s+1}{2j+2}(-1)^{s+\frac12} Y_{j+\frac12,-s-\frac12}\, Y_{j+\frac12,s+\frac12} \\
&\quad - \delta\sqrt{\frac{j-s}{2j}}\sqrt{\frac{j+s+1}{2j+2}}\,(-1)^{s+\frac12}\, Y_{j-\frac12,-s-\frac12}\, Y_{j+\frac12,s+\frac12}\\
&- \delta\sqrt{\frac{j-s}{2j}}\sqrt{\frac{j+s+1}{2j+2}}\,(-1)^{s+\frac12}\, Y_{j+\frac12,-s-\frac12}\, Y_{j-\frac12,s+\frac12} 
\Bigg\} \ .
    \end{split}
\end{equation}
Now that we have all our ingredients, we can focus on the $L=0$ mode in eq. \eqref{eq:EOM_phi_separating2}. In doing so, we observe that the coefficient $K^{(\delta_1,\delta_2)}_{LM\ell_1\ell_2s_1s_2}$ becomes zero for $L=0$, and the double sums in $\ell_{1,2}, \ s_{1,2}$ become single sums. This can be seen either by direct computation using the above expressions, or by noticing that by inserting the expansions in eq.s \eqref{eq:ferm_exp_2} and \eqref{eq:phi_exp_2} in the scalar action $\mathcal{S}_{\text{Yuk}}$ in eq. \eqref{eq:action_separated} we get the integral for the Yukawa interaction term
\begin{equation}
\mathcal{S}_{\text{Yuk}}\supset  \sum_{\delta_1,\delta_2} \sum_{L,\ell_1,\ell_2} \sum_{M,s_1,s_2}\int d^2x\int_{S^2} d^2\Omega \ (Y_{L M}\chi^{(\delta_1)\dagger}_{\ell_1 s_1}\chi^{(\delta_2)}_{\ell_2 s_2})\phi_{L M}\psi^{(\delta_1)\dagger}_{\ell_1 s_1}\psi^{(\delta_2)}_{\ell_2 s_2}.
\end{equation}
Isolating the $L=0$ term in the sum, we observe that $\displaystyle{Y^0_0(\theta,\varphi)=1/\sqrt{4\pi}}$, so that we can exploit the orthogonality condition in eq. \eqref{eq:harm_orth}, which imposes $\ell_1=\ell_2=\ell$, $s_1=s_2=s$ and $\delta_1=\delta_2=\delta$. Keeping also in mind the transformation of the $\psi^{(\delta)}_{\ell s}$ fields in eq. \eqref{eq:psi_transf} and the subsequent transformation of the fermion bilinear in eq. \eqref{eq:ferm_bilin_transf}, we finally obtain eq. \eqref{eq:zero_mode_EOM} for the $L=0$ scalar mode.
The same orthogonality argument fixes the $L=0$ value of the $J$ coefficients. Since $Y_{00}=1/\sqrt{4\pi}$ is constant, the projection of the spinor-harmonic bilinear on the $L=0$ mode is
\begin{equation}
 \int_{S^2} d\Omega\; Y^{0\,*}_{0}\;\chi^{(\delta)\dagger}_{\ell s}\chi^{(\delta)}_{\ell s}
 =\frac{1}{\sqrt{4\pi}}\int_{S^2} d\Omega\;\chi^{(\delta)\dagger}_{\ell s}\chi^{(\delta)}_{\ell s}=\frac{1}{\sqrt{4\pi}}
\end{equation}
for every $\ell$, $s$ and $\delta$, by the normalization \eqref{eq:harm_orth}. Recalling the explicit factor $\delta$ multiplying $(\psi^{(\delta)\dagger}\sigma_3\psi^{(\delta)})$ in eq.~\eqref{eq:ferm_bilin_transf}, the coefficients entering eq.~\eqref{eq:zero_mode_EOM} are therefore
\begin{equation}
\label{eq:J_corrected}
 J^{(\delta)}_{\ell s}=\delta\qquad \text{for all } \ell,\ s\ .
\end{equation}
Coincidentally, this can also be seen by direct computation from eq. \eqref{eq:coeff_terms}, where, plugging the expansion in eq. \eqref{eq:harmonics_exp} and using the Clebsch-Gordan coefficients values
\begin{equation}
    \begin{split}
        \braket{\ell,m;\ell,-m|0,0}=\frac{(-1)^{\ell+m}}{\sqrt{2\ell+1}}
    \end{split}
\end{equation}
we get
\begin{equation}
    \begin{split}
        \chi^{(\delta)\dagger}_{\ell s}\chi^{(\delta)}_{\ell s} &=\frac{1}{2\sqrt{4\pi}}\Bigg\{\frac{j+s}{2j}+\frac{j-s+1}{2j+2}+\frac{j-s}{2j}+\frac{j+s+1}{2j+2}\Bigg\}Y_{00}+\dots=\frac{1}{\sqrt{4\pi}} \ .
    \end{split}
\end{equation}
Applying eq. \eqref{eq:J_coeff} again we get $J^{(\delta)}_{\ell s}=\delta$.\\
We now turn to the task of solving our equation for the $L=0$ mode of the scalar field. Its solution satisfies
\begin{equation}
\label{eq:phi_eq_part}
    \left(\frac{2}{r}\partial_r+\partial^2_r\right)\phi_{\text{part}}=-\frac{\Tilde{y}}{\sqrt{4\pi}}\frac{1}{r^3}\sum_s\sum_{\ell}\sum_{\delta}J^{(\delta)}_{\ell s}( \psi^{(\delta)\dagger}_{\ell s}\sigma_3\psi^{(\delta)}_{\ell s}).
\end{equation}
The solution can finally be written as
\begin{equation}
    \phi_{\text{part}}=\frac{\Tilde{y}}{\sqrt{4\pi}}\sum_{\ell}\sum_s\sum_{\delta}J^{(\delta)}_{\ell s}F^{(\delta)}_{\ell  s}(r)
\end{equation}
with $F^{(\delta)}_{\ell  s}(r)$ the function reported in eq. \eqref{eq:F2_l}.\\

\section{Quadratic fluctuations of the gauge field}
\label{app:A_m_deformations}
In order to find the quantum fluctuations of the gauge field given the spinor-field fluctuations, let us go back to our action
\begin{equation}
    \label{eq:action}
    \mathcal{S}=\frac{1}{e^2}\int d^4 x\left\{-\frac{1}{2}
A_{\mu
}\Box A^{\mu}+i\bar{\Psi}\slashed{D}\Psi+...\right\}\ ,
\end{equation}
where we recast the gauge field kinetic term exploiting the fact that we chose Lorentz gauge ($\partial_{\mu}A^{\mu}=0$) and where we omit the scalar terms and the Wilson line (as they are not relevant for this discussion). Moreover, let us write the gauge field as
\begin{equation}
    A_{\mu}=A_{\mu,\text{cl}}+A_{\mu,\text{quant}} \ ,
\end{equation}
with $A_{\mu,\text{cl}}=g/r$ and $A_{\mu,\text{quant}}$ the quantum fluctuations. Plugging this in eq. \eqref{eq:action}, we obtain in the Lagrangian a term $\displaystyle{(A_{\mu,\text{cl}}+A_{\mu,\text{quant}})(\Box A^{\mu}_{\text{cl}}-Q\delta^3\delta^{\mu}_0)}$ which goes to zero on account of $A_{\mu,\text{cl}}$ solving the EOM in parentheses. We are hence left with the task of solving for the quantum fluctuations, whose action is exactly the one in eq. \eqref{eq:action}.\\
As for the spinor, we decompose the gauge field $A_{\mu,\text{quant}}$ in spherical harmonics\footnote{We are from this point dropping the subscript “quant" in order to avoid a cumbersome notation.} \cite{Higuchi:1986wu,GonzalezLezcano:2023,GonzalezLezcano:2023cuh}:
\begin{equation}
    \label{eq:vector_decomp}
    \begin{split}
        &A_t(t,r,\theta,\varphi) = \sum_{\ell m }A^{(\ell m)}_t(t,r)Y_{\ell m}(\theta,\varphi),\\
        &A_r (t,r,\theta,\varphi)= \sum_{\ell m }A^{(\ell m)}_r(t,r)Y_{\ell m}(\theta,\varphi),\\
        &\begin{pmatrix}
            A_{\theta}\\
            A_{\varphi}
        \end{pmatrix}=\sum_{\ell, m} \left[ a_{\ell m}^{(e)}(t,r)\, \vec{Y}^{(e)}_{\ell m}(\theta, \varphi) + a_{\ell m}^{(o)}(t,r)\, \vec{Y}^{(o)}_{\ell m}(\theta, \varphi) \right],
    \end{split}
\end{equation}
where $\displaystyle{\vec{Y}^{(e)}_{\ell m}(\theta, \varphi) = \nabla_{\mathbb{S}^2} Y_{\ell m}(\theta, \varphi)}$ and $\displaystyle{\vec{Y}^{(o)}_{\ell m}(\theta, \phi) = \vec{r} \times \nabla_{\mathbb{S}^2} Y_{\ell m}(\theta, \phi)}$.\\
We now plug in the action \eqref{eq:action} the expansion \eqref{eq:vector_decomp} and the spinor one 
\begin{equation}
\label{eq:ferm_exp}
    \Psi=\frac{1}{r^{\frac{3}{2}}}\sum_{\ell, s}\sum_{\delta=\pm}\psi^{(\delta)}_{\ell s}(t,r)\otimes \chi^{(\delta)}_{\ell s}(\hat{n}).
\end{equation}
Separating the $\ell=0$ mode for the gauge field (which is the one we will be ultimately interested in), we see that since $Y_{00}=\text{const}$, for this mode we still have $A_{\theta}=A_{\varphi}=0$. We hence obtain
\begin{equation}
    \label{eq:action_recast}
    \mathcal{S}(A)=-\frac{1}{2}\int d^2x \left\{A^{\mu,\ell=0}\Box A^{\ell=0}_{\mu}+\frac{1}{r^3}\sum_{l,s,\delta}\bar{\psi}^{(\delta)}_{l s}\slashed{A}^{\ell=0}\psi^{(\delta)}_{l s}\ \right\} + \dots \ ,
\end{equation}
where we have integrated out the spherical harmonics and the dots are the $\ell\neq0$ modes of $A_{\mu}$, and $\psi^{(\delta)}_{l s}$ are the fermionic quantum fluctuations. This leaves us with the EOMs for the gauge field fluctuations (dropping the superscript $\ell=0$)
\begin{equation}
    \label{eq:A_EOMs}
    \begin{split}
        &\Box A_t=-\frac{1}{r^3}\sum_{\ell, s, \delta}\bar{\psi}^{(\delta)}_{\ell s}\gamma^0 \psi^{(\delta)}_{\ell s},\\
        &\Box A_r=\frac{1}{r^3}\sum_{\ell, s, \delta}\bar{\psi}^{(\delta)}_{\ell s}\gamma^1 \psi^{(\delta)}_{\ell s}.
    \end{split}
\end{equation}
These lead to a field of the same order as the quantum fluctuations of the scalar field $\Phi$ computed in Section \ref{sec:QED_Yukawa}, as the source term is similar. However, we then perform the Weyl map to $AdS_2\times S^2$ (which results in a scaling $\Phi\to r\Phi$ and no scaling for the gauge field) in which we find
\begin{equation}
    \label{eq:dS_bulk_A}
    \delta\mathcal{S}_{\text{bulk}}(A)=\int_{r_0} dt \sqrt{-g}\frac{\delta \mathcal{L}}{\delta \partial
    _rA_{\mu}}\delta A_{\mu}=-g^{r\alpha}g^{\mu\beta}\int_{r_0} dt \sqrt{-g}F_{\alpha\beta}\delta A_{\mu} \ .
\end{equation}
Here, $F_{\mu\nu}$ contains both the classical and quantum contributions. We have checked that the term containing the classical contribution is, however, irrelevant for the analysis, as it can be removed via a boundary term.
Comparing the fully quantum contribution to $\delta \mathcal{S}_{\text{bulk}}[\Phi]$ in eq. \eqref{eq:84}, we checked that the new contributions in \eqref{eq:dS_bulk_A} either vanish faster with $r_0$, making them subleading, or can be absorbed into boundary terms, making them irrelevant for the analysis in Section \ref{sec:new_RG}.

\section{Details on fermionic quartic deformations}
\label{app:Yukawa_details}
This Appendix is dedicated to showing the details of the computation of the RG flow in Section \ref{sec:new_RG}. We resume our discussion from the boundary conditions in eq.s \eqref{eq:cond_a} and \eqref{eq:cond_b} stemming from the requirement $\delta \mathcal{S}_{\text{tot}}=0$, which we rewrite here for convenience:
\begin{equation}
\label{eq:cond_a_app}
    \begin{split}
        \frac{2\nu}{m+\nu}\,\beta \;-\; 2 r_0^{2\nu} F_0 \beta \;-\; 2 F_0 \alpha
\;-\; 2 p_0 \alpha\,\bar\alpha\alpha \;-\; p_0 r_0^{2\nu}\,\alpha\bar\beta\alpha\\
-\; p_0 r_0^{2\nu}\,\alpha\bar\alpha\beta \;-\; p_0 r_0^{2\nu}\,\beta\bar\alpha\alpha
\;+\; \frac{\tilde{y}^2 r_0}{4\pi(m+\nu)^2(1-2\nu)}\, J\alpha\,\big(\bar\beta J \beta\big) \;=\; 0
    \end{split}
\end{equation}
and
\begin{equation}
    \begin{split}
        \frac{2\nu}{m+\nu}\,\bar\beta \;-\; 2 r_0^{2\nu}\,\bar\beta F_0 \;-\; 2\bar\alpha F_0
\;-\; 2 p_0\,\bar\alpha\bar\alpha\alpha \;-\; r_0^{2\nu} p_0\,\bar\alpha\bar\alpha\beta\\
-\; p_0 r_0^{2\nu}\,\alpha\bar\beta\alpha \;-\; p_0 r_0^{2\nu}\,\bar\beta\,\bar\alpha\alpha
\;+\; \frac{\tilde{y}^2 r_0}{4\pi(m+\nu)^2(1-2\nu)}\,(\bar\alpha J)\big(\bar\beta J\beta\big) \;=\; 0 \ .
    \end{split}
\end{equation}
We then rewrite our second equation above as
\begin{equation}
    \begin{split}
       & \sum_J \bar\beta_J
\left[
\frac{2\nu}{m+\nu}\,\mathds{1}_{JK} \;-\; 2 r_0^{2\nu} F_{0\,JK}
\;-\; p_0 r_0^{2\nu}\,\alpha_K \bar\alpha_J
\;+\; \frac{\tilde{y}^2 r_0}{4\pi(m+\nu)^2(1-2\nu)}\,(\alpha J)_K(J\beta)_J
\right]\\
&= \; 2(\bar\alpha F_0)_K \;+\; 2p_0\,\bar\alpha_K(\bar\alpha\alpha) \;+\; r_0^{2\nu} p_0\,\bar\alpha_K(\bar\alpha\beta) \ .
    \end{split}
\end{equation}
Now, let us define the matrix
\begin{equation}
    M = \mathds{1} - \frac{m+\nu}{2\nu}
\left[
2 r_0^{2\nu} F_0 + p_0 r_0^{2\nu}\,\alpha\bar\alpha + p_0 r_0^{2\nu}(\bar\alpha\alpha)\,\mathds{1}
\;-\; \frac{\tilde{y}^2 r_0}{4\pi(m+\nu)^2(1-2\nu)}(J\beta)(\bar\alpha J)
\right]
\end{equation}
and multiply it from the right:
\begin{equation}
    \bar\beta_L \;=\; \frac{m+\nu}{2\nu}\sum_K
\Big[
2(\bar\alpha F_0)_K \;+\; 2p_0(\bar\alpha\alpha)\,\bar\alpha_K \;+\; p_0 r_0^{2\nu}(\bar\alpha\beta)\,\bar\alpha_K
\Big](M^{-1})_{KL}
\end{equation}
so, being $M$ of the form $M=\mathds{1}-X$, we use $M^{-1} = (\mathds{1}-X)^{-1} \simeq \mathds{1} + X$ so that 
\begin{equation}
\begin{split}
    \bar \beta&=\frac{m+\nu}{2\nu}
\left[
2\bar\alpha F_0 + 2p_0(\bar\alpha\alpha)\bar\alpha + p_0 r_0^{2\nu}(\bar\alpha\beta)\bar\alpha
\right]
+ \frac{(m+\nu)^2}{(2\nu)^2}
\Bigg[
4 r_0^{2\nu}\,\bar\alpha F_0^2 \;+\\
&+\; 2 p_0 r_0^{2\nu}(\bar\alpha F_0 \alpha)\bar\alpha
+ 2 p_0 r_0^{2\nu}(\bar\alpha\alpha)(\bar\alpha F_0)
- \frac{\tilde{y}^2 r_0}{2\pi(m+\nu)^2(1-2\nu)}(\bar\alpha F_0 J\beta)(\bar\alpha J) \;+\\
&+\; 4 p_0 r_0^{2\nu}(\bar\alpha\alpha)(\bar\alpha F_0)
- \frac{\tilde{y}^2 r_0}{2\pi(m+\nu)^2(1-2\nu)}(\bar\alpha\alpha)(\bar\alpha J\beta)(\bar\alpha J)
\Bigg] \ +\;\mathcal{O}(p_0^2)\ .
\end{split}
\end{equation}
We can then substitute this into equation \eqref{eq:cond_a_app} to get $\beta$ as a function of $\alpha$. Doing this and truncating to orders linear in $p$ and quadratic in $F$ gives then
\begin{equation}
    \begin{split}
       \beta &= \frac{(m+\nu) r_0^{-2\nu}}{\nu}
\left(\mathds{1} + \frac{m+\nu}{\nu} F\right) F\alpha
\;+\; \frac{m+\nu}{\nu}\, p\, r_0^{1-4\nu}\,\alpha(\bar\alpha\alpha)
\;+\; \frac{(m+\nu)^2}{\nu^2}\, r_0^{-4\nu}\, p\,\alpha(\bar\alpha F\alpha) \;+ \\
&+\; \frac{(m+\nu)^2}{2\nu^2}\, r_0^{1-4\nu}\, p\,(F\alpha)(\bar\alpha\alpha)
\;-\; \frac{\tilde{y}^2\, r_0^{1-4\nu}}{2\pi\nu(m+\nu)(1-2\nu)}\,(J\alpha)(\bar\alpha F J F\alpha) \;+\\
&+\; \frac{(m+\nu)^3}{2\nu^3}\, p\, r_0^{1-4\nu}\,(F\alpha)(\bar\alpha F\alpha) \ ,
    \end{split}
\end{equation}
as claimed in the main body. We now proceed with the request that this boundary condition does not depend on the cut-off $r_0$.\\
Letting
\begin{equation}
    C\coloneqq \frac{(m+\nu) r_0^{-2\nu}}{\nu}
\left(\mathds{1} + \frac{m+\nu}{\nu} F\right) F \ ,
\end{equation}
we get 
\begin{equation}
    \frac{\partial C}{\partial \log \mu} = 0 \qquad \Rightarrow \qquad \beta_F=-2\nu F+2(m+\nu)F\cdot F \ ,
\end{equation}
From the cubic term, we get
\begin{equation}
\label{eq:cond_M_1}
    \begin{split}
       \alpha_b\,\bar\alpha_c\,\alpha_d \,
\frac{\partial}{\partial t}
\left\{ 
\frac{m+\nu}{\nu}\, p\, r_0^{1-4\nu}\,\delta_{ab}\delta_{cd}
\;+\; \frac{(m+\nu)^2}{\nu^2}\, r_0^{1-4\nu}\, p\,\delta_{ab} F_{cd}
\;+\; \frac{(m+\nu)^2}{2\nu^2}\, r_0^{1-4\nu}\, p\, F_{ab}\delta_{cd}
\right.\\
\left.
 +\; \frac{(m+\nu)^3}{2\nu^3}\, p\, r_0^{1-4\nu}\, F_{ab} F_{cd}
\;-\; \frac{\tilde{y}^2\, r_0^{1-4\nu}}{2\pi\nu(m+\nu)(1-2\nu)}\, J_{ab}(FJF)_{cd}
\right\} = 0 \ ,
    \end{split}
\end{equation}
hence:
\begin{equation}
    \frac{\partial}{\partial t}
\left\{
r_0^{1-4\nu}
\left[
p\left(\delta_{ab} + \frac{m+\nu}{\nu}F_{ab}\right)
\left(\delta_{cd} + \frac{m+\nu}{2\nu}F_{cd}\right)
\;-\; \frac{\tilde{y}^2}{2\pi(m+\nu)^2(1-2\nu)}\, J_{ab}(FJF)_{cd}
\right]
\right\} = 0
\end{equation}
\bigskip
\begin{equation}
\label{eq:M}
    \begin{split}
        \Rightarrow\; & -(1-4\nu)
\left[
p\left(\delta_{ab} + \frac{m+\nu}{\nu}F_{ab}\right)
\left(\delta_{cd} + \frac{m+\nu}{2\nu}F_{cd}\right)
\;-\; \frac{\tilde{y}^2}{2\pi(m+\nu)^2(1-2\nu)}\, J_{ab}(FJF)_{cd}
\right] \\
& +\; \beta_p\left(\delta_{ab} + \frac{m+\nu}{\nu}F_{ab}\right)
\left(\delta_{cd} + \frac{m+\nu}{2\nu}F_{cd}\right)
+ p\,\frac{m+\nu}{\nu}
\left[
(\beta_F)_{ab}\left(\delta_{cd} + \frac{m+\nu}{2\nu}F_{cd}\right) +
\right.\\
&\left.
+\; \left(\delta_{ab} + \frac{m+\nu}{\nu}F_{ab}\right)\cdot\frac{1}{2}\,(\beta_F)_{cd}
\right] = 0 \ .
    \end{split}
\end{equation}
One last step is in order: we need to consider Fermi statistics. We need to in fact remember that the $\alpha$'s are Grassmann variables, i.e. $\alpha_a\alpha_b=-\alpha_b\alpha_a$, so that we should actually antisymmetrise what is multiplying these products. Hence, defining
\begin{equation}
    \begin{split}
        & A := \mathds{1} + \frac{m+\nu}{\nu} F
\qquad\qquad
D := FJF \\
& B := \mathds{1} + \frac{m+\nu}{2\nu} F
\qquad\qquad
\lambda := \frac{\tilde{y}^2}{2\pi(m+\nu)^2(1-2\nu)}
    \end{split}
\end{equation}
we can rewrite eq. \eqref{eq:M} as an equation for a $16\times16$ matrix, i.e.
\begin{equation}
    \begin{split}
        \mathcal{M}&\coloneqq  -(1-4\nu)\Big[p\, A\otimes B - \lambda\, J\otimes D\Big]\\
        &+\; \beta_p\, A\otimes B \;+\; p\,\frac{m+\nu}{\nu}
\left(\beta_F \otimes B + \frac{1}{2} A \otimes \beta_F\right) =0 \ .\\
    \end{split}
\end{equation}
Looking at eq. \eqref{eq:cond_M_1}, with the right antisymmetrisation the equation we need to solve in order to find $\beta_p$ is:
\begin{equation}
     \mathcal{M}_{(ac)(bd)} - \mathcal{M}_{(ac)(db)}=0 \ ,
\end{equation}
where our notation means:
\begin{equation}
    (A\otimes B)_{(ac)(bd)} = A_{ab}\, B_{cd} \ .
\end{equation}
Now, to carry out the antisymmetrisation $b \leftrightarrow d$ we introduce a
$16\times16$ matrix $K$:
\begin{equation}
    K\coloneqq \sum_{i,j=1}^4 E_{ij}\otimes E_{ji} \ ,
\end{equation}
with $(E_{ij})_{ab}=\delta_{ia}\delta_{jb}$, so that
\begin{equation}
    K_{(ac)(bd)} = \sum_{i,j=1}^{n} (E_{ij})_{ab}(E_{ji})_{cd}
= \sum_{i,j=1}^{n} \delta_{ia}\delta_{jb}\delta_{jc}\delta_{id}
= \delta_{ad}\delta_{cb} \ .
\end{equation}
Then, for some matrix $M=M_1\otimes M_2$, we have
\begin{equation}
\begin{split}
       (MK)_{(ac)(bd)} &= \Big[(M_1\otimes M_2)\sum_{i,j}E_{ij}\otimes E_{ji}\Big]_{(ac)(bd)}
= \Big[\sum_{i,j=1}^{n} (M_1 E_{ij})\otimes (M_2 E_{ji})\Big]_{(ac)(bd)}\\
& = \sum_{i,j=1}^{n} (M_1 E_{ij})_{ab}\,(M_2 E_{ji})_{cd}
= \sum_{i,j=1}^{n} M_{1,a\alpha}(E_{ij})_{\alpha b}\, M_{2,c\gamma}(E_{ji})_{\gamma d}\\
& = \sum_{i,j=1}^{n} M_{1,a\alpha}\,\delta_{i\alpha}\delta_{jb}\, M_{2,c\gamma}\,\delta_{j\gamma}\delta_{id}
= M_{1,a\alpha}\,\delta_{d\alpha}\, M_{2,c\gamma}\,\delta_{\gamma b}
= M_{1,ad}\,M_{2,cb}
= M_{(ac)(db)}
\end{split}
\end{equation}
as desired\footnote{We observe that our $\mathcal{M}$ above is not a single $M_1\otimes M_2$ matrix, but a rather sum of such terms. However,
\[
MK = \sum_{\alpha,\beta} (M_\alpha \otimes M_\beta)\,K \ ,
\] so that the antisymmetrisation still works for $\mathcal{M}$.}.\\
Finally then our equation is
\begin{equation}
     \mathcal{M}(\mathds{1} - K) = 0 \ .
\end{equation}
Since we only need one information from this matrix, i.e. $\beta_p$, it is sufficient to take the trace of this matrix, so that the equation we actually solve is
\begin{equation}
    \text{Tr}\Big[ \mathcal{M}(\mathds{1} - K)\Big]=0 \ .
\end{equation}
To do this, it is useful to remember a couple of properties:
\begin{equation}
    \operatorname{Tr}(M_1\otimes M_2) = \operatorname{Tr}(M_1)\operatorname{Tr}(M_2) \ ,
\end{equation}
and
\begin{equation}
    \operatorname{Tr}\big((A\otimes B)K\big) = \operatorname{Tr}(AB) \ .
\end{equation}
So, recalling that $\text{Tr}[J] =\text{Tr}\big[ \operatorname{diag}(1,1,-1,-1)\big]=0$, and that, since $J=\Sigma^{q_0}$ anticommutes with $\Sigma^{k_0}$ and $\Sigma^{h_0}$ while commuting with $\Sigma^{f_0}$ and $\Sigma^{q_0}$, one has $JFJ=f\Sigma^{f_0}-k\Sigma^{k_0}-h\Sigma^{h_0}+q\Sigma^{q_0}$ and hence
\begin{equation}
\operatorname{Tr}(JD)=\operatorname{Tr}(JFJF)=4\left(f^2-h^2-k^2+q^2\right), \qquad \operatorname{Tr}(D)=\operatorname{Tr}(JF^2)=4fq \ ,
\end{equation}
our equation becomes:
\begin{equation}
    \begin{split}
        &-(1-4\nu)\Big[p\,\operatorname{Tr}(A)\operatorname{Tr}(B) - \lambda\,\operatorname{Tr}(J)\operatorname{Tr}(D)\Big] +\\
    &    +\; \beta_p\operatorname{Tr}(A)\operatorname{Tr}(B)
+ p\,\frac{m+\nu}{\nu}\operatorname{Tr}(\beta_F)
\left[\operatorname{Tr}(B) + \frac{1}{2}\operatorname{Tr}(A)\right] +\\
&+(1-4\nu)\Big[p\,\operatorname{Tr}(AB) - \lambda\,\operatorname{Tr}(JD)\Big] - \beta_p\operatorname{Tr}(AB) +\\
&-\; p\,\frac{m+\nu}{\nu}
\left\{\operatorname{Tr}(\beta_F B) + \frac{1}{2}\operatorname{Tr}(A\beta_F)\right\} = 0 \ .
    \end{split}
\end{equation}
We now extract $\beta_p$ step by step, to linear order in $p$ and quadratic order in $F$. At this order the overall factor multiplying $\beta_p$ may be evaluated at $F=0$:
\begin{equation}
 \operatorname{Tr}(A)\operatorname{Tr}(B)-\operatorname{Tr}(AB)\;=\;16-4+\mathcal{O}(F)\;=\;12+\mathcal{O}(F)\ .
\end{equation}
The $\lambda$ terms contribute, using $\operatorname{Tr}(J)=0$ and $\operatorname{Tr}(JD)=4(f^2-h^2-k^2+q^2)$,
\begin{equation}
 (1-4\nu)\,\lambda\,\Big[\operatorname{Tr}(J)\operatorname{Tr}(D)-\operatorname{Tr}(JD)\Big]\Big/12
 = -\,\frac{(1-4\nu)\,\tilde y^2\left(f^2-h^2-k^2+q^2\right)}{6\pi(m+\nu)^2(1-2\nu)}\ .
\end{equation}
The terms proportional to $\beta_F$ are kept \emph{unexpanded}, as in eq.~\eqref{eq:delta_beta_run}:
\begin{equation}
 \Delta\beta_p^{\rm run} = -\,p\,\frac{m+\nu}{12\nu}\Big[\operatorname{Tr}\beta_F\Big(\operatorname{Tr}B+\tfrac12\operatorname{Tr}A\Big)-\operatorname{Tr}(\beta_F B)-\tfrac12\operatorname{Tr}(A\beta_F)\Big]\ .
\end{equation}
Being proportional to the full $\beta_F$, this expression vanishes, together with its $p$-derivative, at every fixed point of $F$. On the $h=k=q=0$ slice one finds, exactly,
\begin{equation}
 \Delta\beta_p^{\rm run} = -\,p\,\frac{m+\nu}{2\nu}\Big(3+\frac{2(m+\nu)f}{\nu}\Big)\,\beta_f\ ,\qquad \beta_f=-2\nu f+2(m+\nu)f^2\ ,
\end{equation}
whose leading small-$F$ expansion is $+3(m+\nu)fp$ (using $\beta_F\simeq-2\nu F$, $\operatorname{Tr}\beta_F=-8\nu f$ and $\operatorname{Tr}B+\tfrac12\operatorname{Tr}A\to6$). We stress that the expanded form may not be substituted at the interacting fixed points of $\beta_F$, where $f_*$ is of order one: there, $3(m+\nu)f_*p=3\nu p\neq0$, while the unexpanded expression vanishes because $\beta_f(f_*)=0$. Collecting the pieces we finally find
\begin{equation}
    \beta_p = (1-4\nu)p\;+\;\Delta\beta_p^{\rm run}\;+\; \frac{(4 \nu -1)\, \tilde y^2 \left(f^2-h^2-k^2+q^2\right)}{6 \pi (m+\nu)^2 (2 \nu -1)}
\end{equation}
as promised: this single beta function is the one used for the fixed-point analysis and for the vector field of Fig.~\ref{fig:phasediagram2}. We also note that since at the interacting fixed points $f$ is of order one, the truncation at quadratic order in $F$ (e.g. evaluating the prefactor $\operatorname{Tr}A\operatorname{Tr}B-\operatorname{Tr}AB$ at $F=0$ rather than at $F=F_*$) affects the value of the $\beta_p$ fixed point $p_*$ at $\mathcal{O}(1)$; a treatment exact in $F$ at linear order in $p$ rescales $p_*$ by a factor $\tfrac13$ at the second fixed point, i.e. $p_*^{\text{exact-}F}=\frac{\nu^2\tilde y^2}{18\pi(2\nu-1)(m+\nu)^4}\simeq-0.0108$ at the parameters of Fig.~\ref{fig:phasediagram2}. This truncation ambiguity should be kept in mind when comparing with conformal perturbation theory near $\nu=\tfrac14$.

\end{appendices}

{\fontfamily{bch}\selectfont

\bibliographystyle{unsrt}
\bibliography{refs.bib}

}

\end{document}